\documentclass{WileyMSP-template}
\usepackage{hyperref,braket,graphicx,mathtools,amsthm,physics,algorithm,algcompatible,varwidth,newfloat,xcolor,amssymb}

\newtheorem{mydef}{Definition}

\newtheorem{myprf}{Proof}

\newtheorem{myconj}{Conjecture}

\newcommand{\de}{\textsc{diag-el}}
\newcommand{\des}{\textsc{diag-els}}
\newcommand{\pas}{\textsc{probamps}}
\newcommand{\pa}{\textsc{probamp}}

\newcommand{\oss}{\textsc{optswaps}}
\newcommand{\tg}{\textsc{target}}

\newcommand{\os}{\textsc{optswap}}

\begin{document}

\pagestyle{fancy}
\rhead{\includegraphics[width=2.5cm]{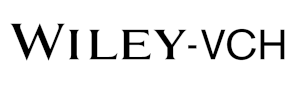}}

\title{Optimal Entropy Compression and Purification in Quantum Bits}

\maketitle

\author{Varad R.~Pande}

\begin{affiliations}
Varad R.~Pande\\
Department of Physics, University of Maryland Baltimore County (UMBC), Baltimore, MD 21250, USA,\\
Department of Physics, Indian Institute of Science Education and Research (IISER) Pune 411008, Maharashtra, India,\\
Harish-Chandra Research Institute~(HRI), HBNI, Jhunsi, Prayagraj (Allahabad) 211019, Uttar Pradesh, India,\\
Skolkovo Institute of Science and Technology (Skoltech), Bolshoy Blvd. 30. 1, Moscow 121205, Russia. 

Email Address: varadrpande@gmail.com

\end{affiliations}


\begin{abstract}

Global unitary transformations (\oss) that optimally increase the bias of any mixed computation qubit in a quantum system – represented by a diagonal density matrix – towards a particular state of the computational basis which, in effect, increases its purity are presented. Quantum circuits that achieve this by implementing the above data compression technique – a generalization of the 3B-Comp
used before – are described. These circuits enable purity increment in the computation qubit by maximally transferring part of its von Neumann or Shannon entropy to any number of surrounding qubits and are valid for the complete range of initial biases. Using the optswaps, a practicable new method that algorithmically achieves hierarchy-dependent cooling of qubits to their respective limits in an engineered quantum register opened to the heat-bath is delineated. In addition to multi-qubit purification and satisfying two of DiVincenzo's criteria for quantum computation in some architectures, the implications of this work for quantum data compression and quantum thermodynamics are discussed.

\end{abstract}

\section{Motivation}
\hspace{0.5cm}
The advent of quantum information theory has resulted in the development of information processing using quantum computers which employ quantum matter and manipulate it according to the rules of quantum mechanics. Certain criteria need to be satisfied for the physical realization of a quantum computer~\cite{divincenzo2000physical}. Two of these are: (i) initialization of computation quantum bits (or qubits) in a well-defined quantum state, and (ii) error correction to tackle environmental decoherence during information processing. Since continuous supply of pure qubits is required for error-correction, the methods relevant to satisfy the former are indirectly necessary for the latter~\cite{knill1997theory}. One such method for qubit initialization is heat-bath algorithmic cooling (HBAC). In algorithmic cooling (quantified using the definition of spin temperature~\cite{PhysRevA.96.012330}), we purify or increase the bias of a required number of qubits to make them available for quantum information processing. At the heart of this procedure lies the transfer of entropy from the computation qubits to the reset qubits followed by exposing the reset qubits to a heat-bath which causes suction of the excess entropy out of them so that the entropy transfer can be repeated.

\hspace{0.5cm}
We first introduce closed-system entropy transfers. The bound~\cite{boykin2002algorithmic} on such transfers can be obtained by considering $n$ qubits, say each with equal bias $ \epsilon $ (or purity $ (1 + \epsilon^2)/2 $) in the computational basis ($ \rho = \{\{(1+\epsilon)/2,0\},\{0, (1-\epsilon)/2\}\} $) and total Shannon entropy $ H_t = nH $, where $ H = [(1 + \epsilon)/2] \log((1 + \epsilon)/2) + [(1 - \epsilon)/2] \log((1 - \epsilon)/2) $. After the (hypothetical) entropy compression, let $m$ qubits be in a pure state so that their Shannon entropy is zero. Consequently, the $n-m$ qubits will each have an entropy $H'$ such that the total is conserved and we have $ (n-m)*H' = n*H $ leading to $ H' = (n/(n-m))*H $. Since the qubits are hotter than before, but not infinitely hot (maximally mixed), we get $ H < H' \leq 1 $, which leads to a bound on the number of qubits that can be completely purified: $ m \leq n(1-H) $. The bound on entropy compression for an $n + 1$ equal-bias system, obtained by Taylor expanding~\cite{atia2016algorithmic} $ H $ to first order in $ \epsilon $ is also equivalent to the one obtained by conservation of purity (or of ``spin order"~\cite{sorensen1989polarization}) and is given by $ \epsilon' = \sqrt{n}*\epsilon $, where $ \epsilon $ and $ \epsilon' $ are initial and final single target qubit biases respectively with the excess entropy being dumped to the $n$ qubits.

\hspace{0.5cm}
Now we turn to a bound tighter than the entropy bound based on fundamental properties of normal matrices. Density matrices representing quantum states are Hermitian, which are a subset of normal matrices and are therefore diagonalizable by a unitary transformation. The purity of a density matrix ($ \Tr(\rho^2) $) does not change under a unitary transformation (the diagonalization). So the ensuing discussion about the possible increase in purity (or bias, once diagonalized) of the target qubit applies most generally. We use the fact that eigenvalues of a Hermitian (more generally, normal) matrix are invariant under any unitary transformation. Therefore, any possible operation to increase the bias of a single target qubit (say) is bound by \textit{exchanges} of diagonal elements (the eigenvalues) of the global (multiqubit) density matrix such that the largest eigenvalues lie in the first half of the resultant density matrix and the smallest ones lie in the bottom half. Using this fact, an analytical bound arises automatically if one arranges all the eigenvalues of the initial state in descending order and require a part of the final state to be such that the purity (or bias) of the target qubit is 1 and the purity of all the rest of the qubits is zero. Let the initial density matrix be $ \rho_i $ and the final density matrix be $ \rho_f = \alpha \rho_{f1} + \beta \rho_{f2} $, where $ \rho_{f1} $ represents part of the Hilbert space which satisfies the aforementioned purity requirement. Then, we have $ \alpha = \Tr(\rho_i \rho_{f1})/\Tr(\rho_{f1}^2) $, and the final bias of the target qubit is given by $ \epsilon' = \alpha - \beta$ (note that $ \beta $ can be obtained from the conservation of trace with respect to $ \rho_i $). This fundamentally inviolable bound was discovered~\cite{sorensen1990universal,sorensen1991entropy} to be smaller than the entropy bound for special systems where there are qubits with only two different biases of the order of $ 10^{-5} $: the $I_n S$~\cite{sorensen1989polarization,sorensen1990universal} or the $ I_n S_m $~\cite{sorensen1991entropy} spin systems. Here, it is also found that the entropy bound violates eigenvalue invariance of normal matrices under unitary transformations and is therefore automatically disproved. While this treatment neatly tells us what the closed system bound could be for these simple spin systems, it does not tell what those purity-maximizing exchanges are or how to implement them. This is done~\cite{sorensen1989polarization} for the special cases in the nuclear magnetic resonance (NMR) architecture up to 7-qubit systems. Star-topology systems in NMR allows to do this for systems as large as 37 qubits~\cite{PhysRevA.96.012330}; in this case several of the diagonal elements bunch together in degenerate energy levels making the exchanges easier.

\hspace{0.5cm}
However, when the default qubit biases (purities) are all unequal, single shot optimal compression is inspired by data compression techniques. In particular, Reference~\cite{schulman1999molecular} used the reverse of von Neumann's method of extracting fair coin flips from a biased coin~\cite{von195113}: apply a C-NOT gate on successive pairs of qubits and keep the control qubit conditioned on the measurement outcome of the NOT qubit being 0, resulting in a boosting of its bias; follow this up by segregation of hot and cold qubits. As such, this is a non-unitary method. A single-shot maximally compressive \textit{unitary} method for purification of the target qubit when three qubits have different biases employs the 3B-Comp~\cite{atia2016algorithmic,baugh2005experimental,brassard2005experimental,elias2006optimal,mor2005algorithmic,park2015hyperfine,park2016heat,kaye2007cooling,ryan2008spin,elias2011heat,zaiser2018experimental,atia2014quantum,mor:2008}, 
a data compression circuit which implements the transformation $ \ket{011} \leftrightarrow \ket{100}  $ in a 3-qubit quantum register. This was introduced and implemented by Reference~\cite{chang2001nmr} and identified in its current form by Reference~\cite{fernandez2004algorithmic} (see Figure 1 in Reference~\cite{chang2001nmr}, Figure 2 in Reference~\cite{fernandez2004algorithmic}, or Figure 2 in Reference~\cite{baugh2005experimental}). Also, the bonacci series of algorithms~\cite{brassard2014prospects,elias2006optimal,elias2007optimal,elias2011semioptimal} perform compression by swapping the bottom element in top half of the global density matrix with the top element in bottom half (see discussion above); the k-bonacci does this for density matrices of k successive qubits while the all-bonacci does it for all qubits below the target. So, the first purpose of this paper is to present optimal entropy-compressive unitary transformations (the aforementioned exchanges, \textit{optswaps}) and the quantum circuits to implement those transformations when the initial biases in any number of qubits could be all different and could take any value between 0 and 1. We shall call the quantum circuits as NB-MaxComp to distinguish them from the ``NB-Comp" unitaries already used by Reference~\cite{elias2006optimal} for the bonacci series of algorithms. We shall provide the circuit LIM-Comp (``LIM" implies a quantum circuit which implements the limiting swap (introduced in section~\ref{limiswap}) which corresponds to the NB-Comp unitaries) for implementing the NB-Comp unitaries as well.

\hspace{0.5cm}
In order to achieve increase in biases beyond the closed-system entropy compression bounds, we need to open the system to a heat-bath which acts as a constant sink of excess entropy by being in contact with some of the qubits involved in the closed system transfer. This was first done by Reference~\cite{boykin2002algorithmic} who used the same primitive as Reference~\cite{schulman1999molecular} and converted it to an open system method -- termed heat-bath algorithmic cooling (HBAC) -- to achieve theoretically better cooling with lesser qubits. The question of finding the ultimate limit of cooling a single qubit with any such open-system method was addressed for the first time by Reference~\cite{schulman2005physical,schulman2007physical}, who proposed the partner pairing algorithm (PPA) to achieve that bound. The PPA consists of a SORT step for closed-system entropy compression which sorts diagonal elements in a descending order and a RESET step for refreshing the hotter qubits by swapping their biases with a RESET-qubit in contact with the heat-bath. Existence of the limit was proven and it was found to lie in the interval $ 2^{n-2} \epsilon \leq \epsilon_f \leq 2^{n-1} \epsilon $, where $ \epsilon_f $ is the final (limiting) bias of the target qubit~\cite{schulman2005physical,schulman2007physical,elias2011semioptimal}. Under the approximation that initial biases $ \epsilon \ll 1/2^n $ ($n$ is total number of qubits participating in the SORT step), the limiting bias of a single qubit was found to be $ 2^{n-2} \epsilon $ for the PPA, and in an independent work, for the all-bonacci algorithm~\cite{elias2006optimal,elias2011semioptimal}.
The former was proven by Reference~\cite{raeisi2015asymptotic} by arriving at an optimal asymptotic state (OAS) which is invariant under PPA. 
The exact HBAC bound was analytically proven using a PPA steady state analogous to the OAS by Reference~\cite{rodriguez2016achievable} by using $m$ reset qubits instead of 1, which reduces to the aforementioned low $ \epsilon $ limit (after substituting $m=1$). 
However, the exact dynamics of HBAC which seek to use optimal intra-subspace entropy transfer have been unknown~\cite{elias2006optimal,elias2011semioptimal,raeisi2015asymptotic,raeisi2019novel}: (i) the PPA dynamics have not been discovered and the transformations needed to implement its sort step have proven elusive; (ii) the bonacci series of algorithms do not account for the loss in biases of the $k - 1$ qubits below the $ k^{th} $ qubit when a kB-Comp is applied; (iii) the method which uses the two-sort operator takes a different route and an unconventional refresh step. So, the second purpose of this paper is to present a HBAC method distinct from these, but one which efficiently -- in minimum total steps and maximum per step compression -- leads to the appropriate exact limits.
This method is operationally systematic in that it leads all the qubits within a quantum register to their respective multi-round limits -- something that gives a high degree of control while purifying the register -- by telling us which unitary transformations to do at what stage of the process. Furthermore, the explicit nature of the code (algorithm) allows us to consider cases where all qubits could have different initial (or default) biases, for example because the qubits see different local environments or are acted upon by a different number of heat-causing quantum operations.

\hspace{0.5cm} 
In section~\ref{GCS}, we shall find the optswaps resulting from a general compression subroutine, conjecture their optimality in section~\ref{GCSopt}, find a numerical proof of these transformations in section~\ref{GCSoptproof}, and quantum circuits for implementing these unitary transformations in section~\ref{NB}. Then we work on open-system compression in section~\ref{Osyst} where we introduce the limiting swap in section~\ref{limiswap}, analytically derive the multi-round limit for hierarchical cooling of a multiqubit quantum register in section~\ref{MRL}, find the algorithm which provides numerical limits supplementing the above and generalizes to the case where the qubits have different default biases in section~\ref{NL}, and build the register compression subroutine (RCS) characterizing the dynamics of HBAC and leading to the multi-round limits in section~\ref{RCS}. Finally, we shall discuss the complexity of RCS-HBAC and the NB-MaxComp in section~\ref{comp} and end with a discussion of this work in section~\ref{disc}.

\section{General Compression Subroutine}
\label{GCS}
\hspace{0.5cm}
Consider an array of qubits in a quantum register. Generally these qubits would be in a mixed state diagonal in some ``natural'' basis. Thus, the state of the $ i^{th} $ computation qubit in the register is, 
\begin{eqnarray}
	\label{singleq}
	\rho_i = 
	\begin{pmatrix*}
		\frac{1+\epsilon_i}{2} && 0 \\
		0 && \frac{1-\epsilon_i}{2}
	\end{pmatrix*}
	=
	\begin{pmatrix*}
		p_i && 0 \\
		0 && 1-p_i
	\end{pmatrix*}
\end{eqnarray} 
where $p_i \in (1/2,1)$ maps to $ \epsilon_i \in (0,1) $ and its Shannon entropy is given by $H_i = (1 + \epsilon_i)/2 \log((1 + \epsilon_i)/2) + (1 - \epsilon_i)/2 \log((1 - \epsilon_i)/2) $. Hereafter, we shall denote $ (1 + \epsilon_i)/2 \equiv \epsilon^+_i $ and $ (1 - \epsilon_i)/2 \equiv \epsilon^-_i $. 
Note that since $ \rho_i $ is diagonal, any tensor product between the density matrices of different qubits, such as $ \rho_1 \otimes \rho_2 \otimes \rho_3 $ (for a three qubit register) will also be diagonal. Since $ \rho_i = \epsilon_i^+ \ket{0}_i\bra{0} + \epsilon_i^- \ket{1}_i\bra{1} $, for the sake of brevity, we can drop the bras and simply write $ \rho_i \equiv \ket{\psi_i} = \epsilon_i^+ \ket{0}_i + \epsilon_i^- \ket{1}_i $ (note that this is \textit{not} a pure state). So, for a three-qubit register with equal biases $ \epsilon $, instead of writing $ \rho_t = (\epsilon^+)^3 \ket{000}\bra{000} + (\epsilon^+)^2 (\epsilon^-) \ket{001}\bra{001} + (\epsilon^+)^2 \epsilon^- \ket{010}\bra{010} + (\epsilon^+) (\epsilon^-)^2 \ket{011}\bra{011} + \epsilon^- (\epsilon^+)^2 \ket{100}\bra{100} + (\epsilon^-)^2 (\epsilon^+) \ket{101}\bra{101} + (\epsilon^-)^2 (\epsilon^+) \ket{110}\bra{110} + (\epsilon^-)^3 \ket{111}\bra{111} $, we write $ \rho_t \equiv \ket{\psi_t} = (\epsilon^+)^3 \ket{000} + (\epsilon^+)^2 (\epsilon^-) \ket{001} + (\epsilon^+)^2 \epsilon^- \ket{010} + (\epsilon^+) (\epsilon^-)^2 \ket{011} + \epsilon^- (\epsilon^+)^2 \ket{100} + (\epsilon^-)^2 (\epsilon^+) \ket{101} + (\epsilon^-)^2 (\epsilon^+) \ket{110} + (\epsilon^-)^3 \ket{111} $. We shall use this notation for rest of the paper. The object of algorithmic cooling is to compress entropy out from some of the qubits in the register to the remaining qubits by increasing the bias ($\epsilon_i$) or probability ($p_i$) of the computation qubits towards state $ \ket{0} $. This is akin to increasing their purity $ \Tr(\rho_i^2) $.

\begin{figure}
	\centering	
	\includegraphics[width=\linewidth]{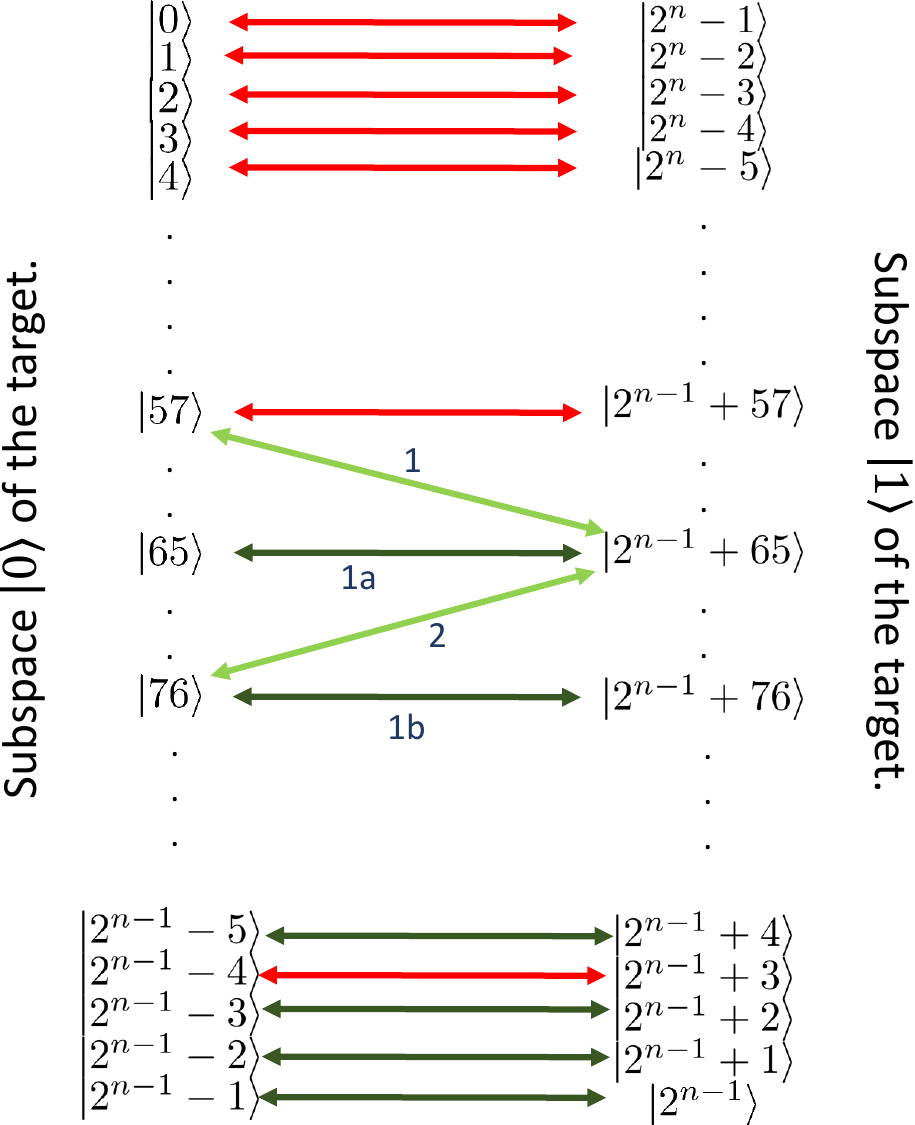}
	\caption{
Graphical representation of Conjecture~\ref{thm1}. We seek to cool the first qubit with its 0 and 1 subspaces shown on the left and right of the figure. Red double-edged arrows represent complementary swaps (section 2.1 in text) between the subspaces which are not beneficial. Dark green double-edged arrows represent the complementary swaps which are beneficial, that is, the \oss. 
Light green arrows 1 and 2 are examples of beneficial non-complementary swaps which are ruled out by statements 1 and 2 of Conjecture~\ref{thm1} respectively with 1a and 1b representing the corresponding beneficial complementary swaps in accordance with Conjecture~\ref{thm1}.}
	\label{OSconjecture}
\end{figure}

\hspace{0.5cm}
For a given array of qubits, we seek to increase the bias of the \textsc{target} qubit towards the state $ \ket{0} $ at the expense of decreasing the corresponding bias of the ancilla qubits. Therefore, the probability amplitudes (\textsc{probamps}) of the state can be divided into two subspaces -- one corresponding to density matrix diagonal elements (\textsc{diag-els}) where the \textsc{target} qubit is $ \ket{0} $ and the other corresponding to the \textsc{diag-els} where the target qubit is $ \ket{1} $. Hereafter, we will call them the $ \ket{0T} $ and the $ \ket{1T} $ subspaces respectively (see pictorial representation in Figure~\ref{OSconjecture}). Thus, the system of $n$ qubits, each represented by Equation~\ref{singleq}, can be expressed as:
\begin{eqnarray}
	\label{system}
	\ket{\psi_{t}} = \bigotimes_{i=1}^{n} \ket{\psi_i} = \sum_{j = 0}^{2^n - 1} \prod_{\{k_j\}} \prod_{\{l_j\}} \epsilon_{k_j}^+ \epsilon_{l_j}^- \ket{ j }, 
\end{eqnarray}
where $ j $ is the decimal number corresponding to the respective \textsc{diag-els}. $k_j$ and $l_j$ index $0's$ and $1's$ respectively in the n-bit binary equivalent of $j$. Thus, $ k_j, l_j \in \{1,...,n\} $, $ \{ k_j \} \cup \{ l_j \} = \{ 1,...,n \} $ and $ \{ k_j \} \cap \{ l_j \} = \emptyset $. 
To reiterate our notation: for the sake of brevity, in the above and subsequent equations, we set $ \ket{ j }\bra{j} \equiv \ket{ j } $, $ \rho_t \equiv \ket{\psi_{t}} $, and $ \rho_i \equiv \ket{\psi_{i}} $. Without loss of generality, we choose to cool the first qubit. The subspace division can be expressed as:
\begin{eqnarray}
	\label{subspace}
	\ket{\psi_{0T}} &=& \sum_{j = 0}^{2^{n-1} - 1} \epsilon_1^+ \prod_{\{ k_j \}} \prod_{\{ l_j \}} \epsilon_{k_j}^+ \epsilon_{l_j}^- \ket{ j }, \nonumber\\
	\ket{\psi_{1T}} &=& \sum_{j = 2^{n-1}}^{2^n - 1} \epsilon_1^- \prod_{\{ k_j \}} \prod_{\{ l_j \}} \epsilon_{k_j}^+ \epsilon_{l_j}^- \ket{ j }. \nonumber
\end{eqnarray}
The exchange of particular \textsc{probamps} between these two subspaces using certain entropy compressive unitary transformations (\textsc{optswaps}) is the building block of the subroutine. The \textsc{optswaps} conform to the following prescription:
\begin{mydef}
	\label{def1}
	Let the set of unitary transformations $ J \equiv \{ V_j \} $, where $ V_j \equiv \ket{ j }_{0T} \leftrightarrow \ket{ 2^n - 1 - j }_{1T} $ $\forall$ $j \in \{0,..., 2^{n-1} - 1 \}$. Then let us define $ J_{\theta} \subseteq J $ such that, $ V_j \in J_{\theta} $ if and only if $ R_{j,0T} < R_{j,1T} $, where $ R_{j,0T} \equiv (\epsilon_1^+ \prod_{\{ k_j \}} \prod_{\{ l_j \}} \epsilon_{k_j}^+ \epsilon_{l_j}^- )_{0T} $ and $ R_{j,1T} \equiv ( \epsilon_1^- \prod_{\{ k_j \}} \prod_{\{ l_j \}} \epsilon_{k_j}^- \epsilon_{l_j}^+ )_{1T} $.
\end{mydef}
For example, in a 5-qubit system, if $ \epsilon_1^+ \epsilon_2^- \epsilon_3^+ \epsilon_4^- \epsilon_5^- < \epsilon_1^- \epsilon_2^+ \epsilon_3^- \epsilon_4^+ \epsilon_5^+ $, then the corresponding \textsc{optswap} is $ \ket{0_1 1_2 0_3 1_4 1_5} \leftrightarrow \ket{1_1 0_2 1_3 0_4 0_5} $ or simply $ \ket{01011} \leftrightarrow \ket{10100} $. In this case, $ \{k_j\} = \{2,4,5\} $ and $ \{l_j\} = \{1,3\} $. In terms of decimals it can be simply represented as: $ \ket{11} \leftrightarrow \ket{20} $. Upon performing all the \textsc{optswaps} that satisfy the above prescription and denoted by the set $ J_\theta = \{ j_c \} $, the increase in bias of the \textsc{target} is given by $ X_n = \epsilon_1' - 
\epsilon_1 = 2 \sum_{\{j_c\}} ( R_{j_c,1T} - R_{j_c,0T} ) $
\begin{eqnarray}
	&=& 2 \sum_{\{j_c\}} \bigg ( \frac{1}{2^n} \sum_{m} \sum_{A_m, B_m} (-1)^{\sum_{B_m} k_{j_c m} \bmod{(k_{j_c m} - 1)}} \nonumber\\
	&& \prod_{A_m} \epsilon_{l_{j_c m}} \prod_{B_m} \epsilon_{k_{j_c m}}  \bigg),
\end{eqnarray}
where, $ A_m \equiv \{ l_{j_c m} \} $, $ B_m \equiv \{ k_{j_c m} \} $, and $ m =  \mathbf{card} (A_m \cup B_m) \in 2\mathbb{N} + 1 $. Also, $ l_{j_c m}, k_{j_c m} \in \mathcal{C} (A_m \cup B_m ) $, where $ \mathcal{C} $ denotes the set of all possible combinations of elements in $ A_m \cup B_m $. This expression is derived by making empirical observations for some examples.

\subsection{Optimality} 
\label{GCSopt}
\hspace{0.5cm}
The optimality of the \textsc{optswaps} for increasing the bias of the \textsc{target} is based on Definition~\ref{def1}. 
The following conjecture is based on ruling out all the non-complementary swaps (those which are violative of Definition~\ref{def1}) by establishing the fact that they are suboptimal for our task:

\begin{myconj}
	\label{thm1}
	If 
	\begin{enumerate}
		\item $ R_{a,1T} - R_{a,0T} \geq S_{b,1T} - R_{a,0T} $ $ \forall $ $ a \in J_\theta $ and $ \forall $ $ b \in J \setminus J_\theta $, and
		\item given $ R = R_{a,1T} - R_{a,0T} > 0 $ and $ S = S_{b,1T} - S_{b,0T} > 0 $, $ S_{b,1T} - R_{a,0T} < R + S $,  $ \forall $ $ a,b \in J_\theta $ such that $ a \neq b $, and			
		\item given $ J_\theta = \emptyset $, $ S_{b,1T} - R_{a,0T} < 0 $, $ \forall $ $ a,b \in J $,			
	\end{enumerate} 
	then, the increase in bias of the target, $ X_n $, is maximal. 	
\end{myconj}

where $ R_{a,1T} \equiv (\epsilon_1^- \prod_{\{ l_a \}} \prod_{\{ k_a \}} \epsilon_{k_a}^- \epsilon_{l_a}^+)_{1T} $, $ R_{a,0T} \equiv (\epsilon_1^+ \prod_{\{ l_a \}} \prod_{\{ k_a \}} \epsilon_{k_a}^+ \epsilon_{l_a}^-)_{0T} $, $ S_{b,0T} \equiv (\epsilon_1^+ \prod_{\{ l_b \}} \prod_{\{ k_b \}} \epsilon_{k_b}^+ \epsilon_{l_b}^-)_{0T} $ and $ S_{b,1T} \equiv (\epsilon_1^- \prod_{\{ l_b \}} \prod_{\{ k_b \}} \epsilon_{k_b}^- \epsilon_{l_b}^+)_{1T} $. 
We note that for case 3, $ X_n = 0 $. Figure~\ref{OSconjecture} provides a graphical visualization of this conjecture.

\hspace{0.5cm}
	
The complementary swaps are such that the number of zeroes in the subspace $ \ket{0T} $ is the same as number of ones in the $ \ket{1T} $ subspace and vice-versa. Complementary but non-beneficial swaps are those which do not contribute to $ X_n $, that is do not increase the bias of the first qubit. The non-complementary swaps are such that this number is unequal. These swaps are beneficial when a corresponding ket in the $ \ket{1T} $ subspace has a larger value than one in the $ \ket{0T} $ subspace. As an example, for an equal bias 3-qubit system with $ \epsilon $ as the bias of each qubit, the beneficial complementary swap is $ \ket{011} \leftrightarrow \ket{100} $. After implementing this using the 3B-Comp quantum circuit mentioned in the introduction, the sum of $ \pas $ in the $ \ket{0T} $ subspace of the first qubit is increased and that in the $ \ket{1T} $ subspace is decreased. So, the final bias of the first qubit is $ (\epsilon_+^3 + 3 \epsilon_+^2 \epsilon_-) - (\epsilon_-^3 + 3 \epsilon_-^2  \epsilon_+) = 3 \epsilon/2 - \epsilon^3/2 $. This matches with the value found in the literature~\cite{baugh2005experimental,brassard2005experimental,chang2001nmr,elias2006optimal,elias2007optimal,elias2011heat,elias2011semioptimal,fernandez2004algorithmic,rodriguez2016achievable,rodriguez2017heat,mor2005algorithmic,brassard2014prospects,park2015hyperfine,park2016heat,ryan2008spin,schulman2005physical,schulman2007physical,zaiser2018experimental}	
\subsection{Numerical proof}
\label{GCSoptproof}
\hspace{0.5cm}
Numerical proof of the conjecture is obtained through the following pseudocode. With the biases of all qubits in the register as input, it outputs the exact swaps that need to be performed, and verifies the optimality of these swaps by ruling out all other swaps by demonstrating cases 1 and 2 presented in 
conjecture~\ref{thm1}.
As a consequence of Algorithm~\ref{alg1}, all the \pas $ $ in $ \ket{0T} $ subspace would be greater than or equal to all the \pas $ $ in the $ \ket{1T} $ subspace of the $\tg$ qubit. Line 4 of Algorithm~\ref{alg1} describes the $\des$ and $\pas$, line 11 implements Definition~\ref{def1}, and lines 21, 30 and 41 initiate numerical proof statements 1, 2, and 3 of 
conjecture~\ref{thm1} respectively. The conjecture has been demonstrated for several combinations of initial biases and number of qubits (see Appendix). 
\begin{algorithm}[H]
	\caption{\textbf{Optswaps}}
	\label{alg1}
	\begin{algorithmic}[1]
		\REQUIRE Starting biases of the qubits: $ \{ \epsilon_i \}$
		\ENSURE Cool the $m^{th}$ qubit in the register.
		\STATE Bring the bias of the $ m^{th} $ qubit to position 1: 
		swap($\epsilon_1, \epsilon_m$)
		\FOR{$j \leftarrow 0$ to $2^n - 1$}
		\STATE 
		\begin{varwidth}[t]{\linewidth} 
			$\rhd$ $ \de_{j} \leftarrow $ $ (j)_{10} $ converted to its binary equivalent $ (j)_{2} $ 
			composed of $ n $ binary digits by appropriately padding $ 0's $ to the left.
		\end{varwidth}	
		\FOR{$i \leftarrow 1$ to $n$}
		\STATE $\rhd$ $ \epsilon_i^+ \leftarrow (1 + \epsilon_i)/2 $ and $ \epsilon_i^- \leftarrow (1 - \epsilon_i)/2 $.
		\STATE $\rhd$ $\pas_{ij}$ $ \leftarrow $ Assigning $ \epsilon_i^+ $ for $ 0_i $ and $ \epsilon_i^- $ for $ 1_i $ in $ \de_{j} $.
		\ENDFOR 
		\STATE $\rhd$ $\pa_{j}$ $ \leftarrow $ $\prod_{i} \pas_{ij}$
		\ENDFOR
		\STATE The number of swaps $ n_s $ $\leftarrow$ 0.
		\FOR{$ k \leftarrow 0 $ to $2^{n-1} - 1$}
		\IF{$\pa_{k} < \pa_{2^n - k - 1}$}
		\STATE $ \rhd $ swap($\de_{k} , \de_{2^n - k - 1} $). 
		\STATE $ \rhd $ $ n_s \leftarrow n_s + 1 $. 
		\STATE $ \rhd $ swap($ \pa_k, \pa_{2^n - k - 1} $) 
		\STATE $ \rhd $ $\mathbf{Print}$: Swapped $ \ket{k} $ and $ \ket{2^n - 1 - k} $ or $ \de_{k} $ and $ \de_{2^n - k - 1} $.
		\ENDIF 
		\ENDFOR
		\STATE $ \rhd $ Final bias $\epsilon_m' \leftarrow \sum\limits_{j=0}^{2^{n-1} - 1} \pa_{j} - $ $ \sum\limits_{j=2^{n-1}}^{2^n - 1} \pa_{j} $  
		\IF{$n_s \neq 0$}
		\FOR{$ k,l \leftarrow 0 $ to $2^{n-1} - 1$} 
		\IF{$\pa_{k} < \pa_{2^n - k - 1}$ and $ \pa_{l} > \pa_{2^n - l - 1} $}
		\STATE $ \rhd $ $v$ $\leftarrow$ $\pa_{2^n - k - 1} - \pa_{k} $
		\IF{ $ \pa_{2^n - l - 1} - \pa_{k} > v $ }
		\STATE $\mathbf{Print:}$ Exception found; case 1. disproved; 
		swaps are not optimal.
		\ELSE 
		\STATE $\mathbf{Print:}$ Exception not found; case 1. 
		demonstrated. 
		\ENDIF 
		\ENDIF
		\IF{$\pa_{k} < \pa_{2^n - k - 1}$ and $ \pa_{l} < \pa_{2^n - l - 1} $}
		\STATE $\rhd$ $v_1$ $\leftarrow$ $\pa_{2^n - k - 1} - \pa_{k} $
		\STATE $\rhd$ $v_2$ $\leftarrow$ $\pa_{2^n - l - 1} - \pa_{l} $
		\IF{ $ \pa_{2^n - l - 1} - \pa_{k} > v_1 + v_2 $ }
		\STATE $\mathbf{Print:}$ Exception found; case 2. disproved; 
		swaps are not optimal.
		\ELSE 
		\STATE $\mathbf{Print:}$ Exception not found; case 2. 
		demonstrated. 
		\ENDIF 
		\ENDIF		
		\ENDFOR
		\ELSIF{$ n_s $ == 0}
		\FOR{$ k,l \leftarrow 0 $ to $2^{n-1} - 1$} 
		\IF{$ \pa_k < \pa_{2^n - l - 1} $}
		\STATE $\mathbf{Print:}$ Exception found; case 3. disproved; 
		Swaps are not optimal.
		\ELSE
		\STATE $\mathbf{Print:}$ The qubit cannot be cooled. 
		\ENDIF
		\ENDFOR
		\ENDIF
	\end{algorithmic}
\end{algorithm}	
\section{NB-MaxComp}
\label{NB} 
\begin{figure}
	\centering
	\includegraphics[width=\linewidth]{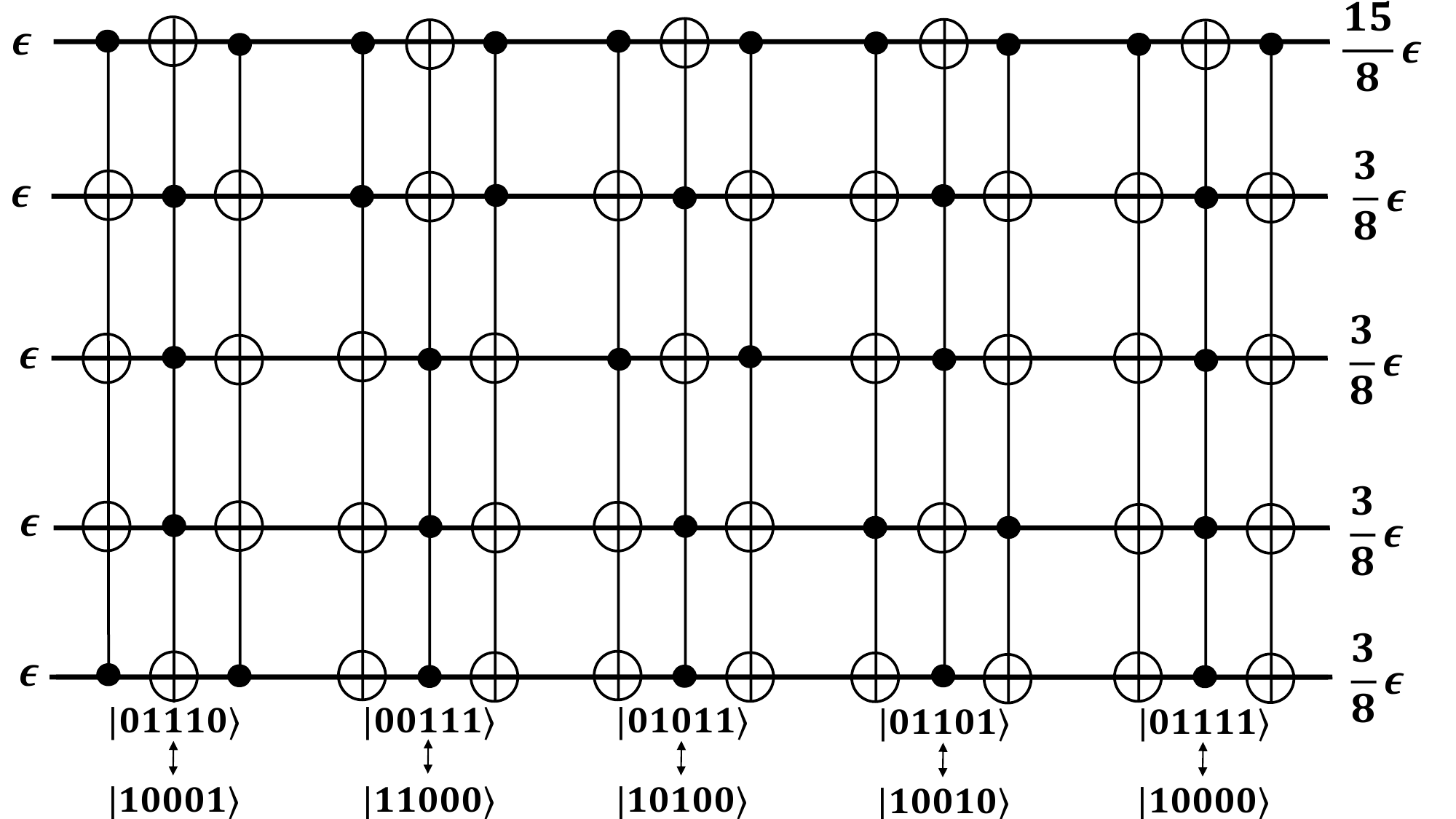}
	\caption{An example of NB-MaxComp: the 5B-MaxComp compression circuit. The \os $ $ corresponding to each unitary transformation is shown below the 3-line gate set which implements it. There are five such sets in this case.}
	\label{5BMAXCOMP}
\end{figure}

\hspace{0.5cm} 
The quantum circuits that implement the unitary transformations proposed above are multi-qubit analogs of the \textit{C-NOT} or \textit{Toffoli}~\cite{nielsen2002quantum} gates. For each swapped \textsc{diag-el} in the $ 0T $ subspace, we put a \textsc{control} gate for each individual swap $ \ket{0}_{OT} \leftrightarrow \ket{1}_{1T} $ and a \textsc{not} gate for each $\ket{1}_{0T} \leftrightarrow \ket{0}_{1T} $. The circuit corresponding to optimal entropy compression in a 5 qubit register with equal initial biases is shown in Figure~\ref{5BMAXCOMP}.

\section{Open-system compression}
\label{Osyst}
\hspace{0.5cm} 
Upon implementing the \oss $ $ on a set of qubits, bias of the first qubit increases and is compensated by the decrease in bias for rest of the qubits because entropy is conserved in the closed system. For example, see the 5-qubit register in Figure~\ref{5BMAXCOMP}. In order to again cool the first qubit, we need to bring the bias in rest of the qubits back to their default or initial (terms used interchangeably in the text) value by bringing the register in contact with an environment which acts as a heat-bath and therefore as an entropy sink. This can be done by surrounding all the qubits with satellite qubits like in a star-topology register~\cite{PhysRevA.96.012330}, or by swapping their biases with a single~\cite{schulman2005physical,raeisi2015asymptotic} or multiple~\cite{rodriguez2016achievable} refrigerant qubits whose sole purpose is to serve as an intermediary between the bath and computation qubits which are sought to be cooled. 

\subsection{Limiting Swap}
\label{limiswap}
\hspace{0.5cm}
To find the limit of cooling a single qubit given a set of ancilla qubits to which its entropy can be transferred, we need to find the \os $ $ which is \textit{last} beneficial \os. This \os $ $, termed as the limiting swap is given by $ \ket{011...1}_{0T} \leftrightarrow \ket{100...0}_{1T} $. It is formalized below:
\begin{myprf}
	\label{limswap}
	Given that $a \in \{ 0,..., 2^{n-1} - 2 \} $ and $ b \in \{2^{n-1} + 1,..., 2^{n} - 1 \} $, $\forall$ $a$ and $b$, we have $R_{a,0T} > R_{2^{n-1} - 1,0T} $ and $R_{b,1T} < R_{2^{n-1},1T} $. 
	It implies that, $ R_{2^{n-1},1T} \leq R_{2^{n-1} - 1,0T} $ if and only if, $\forall$ $a$ and $b$, $R_{a,0T} \geq R_{b,1T} $ which violates Definition~\ref{def1}.  
\end{myprf}
The proof establishes that the limiting swap $ \ket{011...1}_{0T} \leftrightarrow \ket{100...0}_{1T} $ is counterproductive if and only if all other swaps are also counterproductive making it the last productive swap. It can be implemented using the data compression circuit LIM-Comp shown in Figure~\ref{LIMCOMP}.
Bias of the \tg $ $ qubit at the limit can be found by equating the $\pas$ corresponding to the limiting swap: $ (\epsilon_l)_+ \epsilon_-^{m} = (\epsilon_l)_- \epsilon_+^{m} $ ($m$ is the number of reset qubits), which gives us:
\begin{equation}
	\label{frl}
	\epsilon_l = \frac{(1 + \epsilon)^{m} - (1 - \epsilon)^{m}}{(1 + \epsilon)^{m} + (1 - \epsilon)^{m}},
\end{equation}
assuming that default bias of all the ancilla qubits is identically $ \epsilon $ and bias of the computation \tg $ $ qubit is denoted by $ \epsilon_l $. In the example of a star-topology register mentioned above, this limiting bias can act as default bias of the computation qubits in the idealized scenario of infinite relaxation time for the computation qubits. As mentioned, in other cases the default bias would be $ m\epsilon $ for multiple refrigerant qubits or $ \epsilon $ for a single refrigerant qubit.

\subsection{Multi-round limit -- Analytical Proof}
\label{MRL}
\hspace{0.5cm}
Above, we found the limit of cooling a single qubit to its limit assuming that the initial biases of the rest of the qubits are all identically $ \epsilon $. Here, our purpose is to cool a register of multiple qubits where all qubits are cooled to their respective limits. We begin by cooling the first qubit to its limit by utilizing the minimum bias of the rest of the qubits at each compression step. We continue cooling each remaining qubit in the register with the help of respective number of qubits below it, i.e., our subspace would contain one less qubit as we go down the hierarchy. At the end of this procedure, all the qubits in the register are cooled to their first-round limits. Using the first-round limits of all the qubits in the register, we again cool the first qubit, this time to its second-round limit. Again, we proceed to cool each remaining qubit to its respective second-round limit, where the subspace being utilized sees a reduction of one qubit as we go down the hierarchy. We continue this procedure to obtain the multi-round limit of cooling the quantum register.
Thus, the expression for the limit of purifying the $ k^{th} $ qubit in the $ r^{th} $ limiting round in a register of size $n$ is given by equating the $\pas$ corresponding to the limiting swap in respective rounds (similar to the Equation~\ref{frl}):
\begin{equation}
	\label{limit}
	\epsilon_{rkn} = \frac{(1 + \epsilon)^{f(r,k,n)} - (1 - \epsilon)^{f(r,k,n)}}{(1 + \epsilon)^{f(r,k,n)} + (1 - \epsilon)^{f(r,k,n)}},
\end{equation}
where the function $ f(r,k,n) $ is given by recursive relations which we shall derive here. Without loss of generality about the specific refrigerant qubit scenario, we assume the default bias to be $ \epsilon $.

\hspace{0.5cm}
Consider the first qubit in the first limiting round: it is purified to its limit by transferring entropy to the $ n - 1 $ qubits lower in the hierarchy (see limiting swap \ref{limswap}): $(\epsilon_{11n})_+ \epsilon_-^{n-1} = (\epsilon_{11n})_- \epsilon_+^{n-1} $. We thus have $ f(1,1,n) = n-1 $. The second qubit would be purified using only the $ n - 2 $ qubits lower than itself: $ f(1,2,n) = n-2 $. For $ k^{th} $ qubit, we thus have $f(1,k,n) = n - k$. For the last and the penultimate qubit, the function equals just 1. The second round limit for the $ k^{th} $ qubit is obtained by using the first round limits of all the qubits lower in the hierarchy: $[(\epsilon_{2kn})_+] \prod_{i=k+1}^{n-2} \epsilon_-^{n-i} \epsilon_-^2 = [(\epsilon_{2kn})_-] \prod_{i=k+1}^{n-2} \epsilon_+^{n-i} \epsilon_+^2 $. Solving this, we find $ f(2,k,n) = \sum_{i=k+1}^{n-2} f(1,i,n) + 2 $, where the 2 is added for the last two qubits. The third round limit is obtained by adding the function for the second round up to the $ (n-3)^{rd} $ qubit which is added to the function for the first round: $f(3,k,n) = \sum_{i=k+1}^{n-3} f(2,i,n) + f(1,n-2,n) + 2 $. Similarly for the $4^{th}$ qubit, we have $ f(4,k,n) = \sum_{i=k+1}^{n-4} f(3,i,n) + f(2,n-3,n) + f(1,n-2,n) + 2 $, and for the $ 5^{th} $ qubit, we get $ f(5,k,n) = \sum_{i=k+1}^{n-5} f(4,i,n) + f(3,n-4,n) + f(2,n-3,n) + f(1,n-2,n) + 2 $ and so on. 

\hspace{0.5cm}
We notice that as one goes further into the limiting rounds, the limit up to only the $(n - r - 1)^{th}$ qubit shows an increment. Also, we note the difference in expressions of $ f $ corresponding to the $ 1^{st} $ and $ 2^{nd} $ rounds, and that of round 3 and further. Based on this observation, the recursive relation corresponding to $ r > 2 \text{ \& } k < n - r $ is given by: 
\begin{equation}
	\label{f1}
	f(r,k,n) = \sum_{j=1}^{r-2} f(j, n-j-1, n) + \sum_{i=k+1}^{n-r} f(r-1,i,n) + 2.
\end{equation}
Similar observation for $ r = 2 \text{ \& } k < n - r $ yields 
\begin{equation}
	\label{f2}
	f(r,k,n) = \sum_{i=k+1}^{n-r} f(r-1,i,n) + 2.
\end{equation}
Finally, for $ r \geq 2 \text{ \& } k \geq n - r $, we get
\begin{equation}
	\label{f3}
	f(r,k,n) = f(r-1,k,n).
\end{equation}
The initial condition for $ f $ corresponds to the first-round limit of the respective qubits, where $k < n - 1$:
$
f(1,k,n) = n - k$. For $ k \geq n-1 $, we simply have $ f(1,k,n) = 1 $. We also note that for a given $n$, $ r_{max} = n - 2 $. In the low initial bias case, Equation~\ref{limit} can be expanded to first order in $ \epsilon $ for the case $ k = 1, r = r_{max} $, to obtain $ \epsilon_{rkn} = 2^{n-2} \epsilon $, which is consistent with Reference~\cite{elias2011semioptimal,rodriguez2016achievable,raeisi2015asymptotic,schulman2005physical}.

\hspace{0.5cm} 
From Equation~\ref{limit}, together with Equation~\ref{f1}, \ref{f2} or \ref{f3}, we can find the limit of cooling a particular qubit or a particular set of qubits in the quantum register. Thus our quantum information processor achieves flexibility due to its ability to separate the computation space of any size (dependent on the nature of the computation) from the qubits that are just meant to cool the computation qubits. Further, the formula can be used to tailor our needs by fixing two or three of the four variables, $ \epsilon_{rkn} $, $ r $, $ k $, and $ n $ to obtain a space of the remaining variables satisfying the chosen constraint. In Figure~\ref{Limitsvsn}, we show the limits for different values of initial biases when $ r = r_{max} $ and $ k = 1 $. 
In Figure~\ref{Limitsvsrlowe}, we show the change in limits with different limiting rounds when the initial bias is fixed at $ \epsilon = 0.00001 $ for several values of $ n $.
We can also plot (see Figure~\ref{fig5}) the limits with respect to the number of rounds.
\subsection{Numerical Limits}
\label{NL}
\hspace{0.5cm} 
The afore-derived limits are obtained under the assumption that the initial or default biases of all the qubits in the register are equal ($\epsilon$) to begin with. However, to make a statement about the limiting entropy distribution in a quantum register with different default biases (say, due to connection with heat baths at different temperatures), we construct the pseudocode Limits~\ref{alg2}.
As expected, the afore-derived limits can be numerically obtained by implementing this pseudocode. 

\hspace{0.5cm}
Within a given limiting round ($r$), the \textit{for} loop (line 8) truncates the subspace of the qubit register from the $ v^{th} $ qubit to the $ n-r-1^{th} $ qubit. The upper limit on the subspace size is imposed by noting that within a given limiting round, the qubit index $k \leq n - r - 1$. Line 10 enters a \textit{while} loop which repeatedly implements compression using the \oss $ $ till the point where the bias/purity of the $ v^{th} $ qubit can no longer be increased. The \textit{while} loop terminates when the ratio of purities (line 31) before and after compression (single iteration of the while loop) asymptotically reaches 1. When the purity of the $ v^{th} $ qubit reaches its limit within a given round (line 33), it is disconnected from the subspace by the next -- $ v+1^{th} $ -- iteration of the for loop. Further, when the biases/purities of all the qubits reach their limits within a given limiting round, they serve as initial biases of the register for the next -- $ r+1^{th} $ -- round (lines 35 through 39).
 
\hspace{0.5cm} Let us call the first qubit within a compression step as the target and the remaining qubits, lower down in the register, as ancilla (note that this characterization is valid only \textit{within} a given compression step). It should be noted that although we are arriving at the respective limits in this pseudocode within a given compression step (the while loop), we do not account for the purity/bias decrease of the ancilla when we increase the bias of the target qubit by transferring its entropy to the remaining qubits within the compression subspace. This ignorance can be justified with the argument that, within a given round (say $r$), when an ancilla qubit's purity decreases than its limit in the previous round ($ r - 1 $), one can do a series of compression rounds to bring it back to the level of the previous round ($ r - 1 $) using the qubits lower in the hierarchy with respect to the particular ancilla qubit before proceeding to increase the purity of the target within the compression step in the $ r^{th} $ round. The same observation holds for all the ancilla qubits within the compression subspace. This argument is bolstered in the register compression algorithm (see next section) in which the pseudocode Limits serves as a subroutine providing target purities/biases of the register for each limiting round.    

\begin{algorithm}[H]
	\caption{\textbf{Limits}}
	\label{alg2}
	\begin{algorithmic}[1]
		\REQUIRE Number of qubits, $ n $.
		\REQUIRE Starting biases of the qubits, $ \{ \epsilon^{in}_i \} $.
		\REQUIRE Number of \textit{rounds}; highest possible is $ n - 2 $.
		\REQUIRE Desired \textit{precision}.
		\ENSURE Sequentially arrive at the cooling limit of all the qubits in the register.
		\STATE $\rhd$ $\omega_{i} \leftarrow \epsilon^{in}_{i} $
		\FOR{$r \leftarrow 1$ to \textit{rounds}}
		\IF{r $>$ 1}
		\STATE $ \{ \epsilon^{in}_i \} $  $\leftarrow$ $ \{ \epsilon^{f}_{r-1,i} \} $
		\ELSE
		\STATE $ \{ \epsilon^{in}_i \} $ $\leftarrow$ $ \{ \omega_1, \epsilon^{in}_{2:n} \} $ 
		\ENDIF
		\FOR{$v$ $ \leftarrow $ 1 to $ n - r - 1 $}
		\STATE s $\leftarrow$ 0
		\WHILE{Purity increase ratio $\neq$ 1 within the given precision requirement.}
		\STATE $s \leftarrow s + 1$ 
		\IF{$s > 1$}
		\STATE $ \{ \epsilon^{transient}_i \} $ $\leftarrow$ $ \{ \epsilon^{increase}, \epsilon^{in}_{v+1}, ... , \epsilon^{in}_n \} $
		\ENDIF 
		\FOR{$j \leftarrow 0$ to $2^{n-v+1} - 1$}
		\STATE 
		\begin{varwidth}[t]{\linewidth} 
			$\rhd$ $ \de_{j} \leftarrow $ $ (j)_{10} $ converted to its binary equivalent $ (j)_{2} $ composed of $ n $ binary digits by appropriately padding $ 0's $ to the left.
		\end{varwidth}
		\FOR{$i \leftarrow 1$ to $n-v+1$}
		\STATE $\rhd$ $ \epsilon_i^+ \leftarrow (1 + \epsilon_i)/2 $ and $ \epsilon_i^- \leftarrow (1 - \epsilon_i)/2 $.
		\STATE $\rhd$ $\pa_{ij}$ $ \leftarrow $ Assigning $ \epsilon_i^+ $ for $ 0_i $ and $ \epsilon_i^- $ for $ 1_i $ in $ \de_{j} $.
		\ENDFOR 
		\STATE $\rhd$ $\pa_{j}$ $ \leftarrow $ $\prod_{i} \pas_{ij}$
		\ENDFOR
		\FOR{$ k \leftarrow 0 $ to $2^{n-v} - 1$}
		\IF{$\pa_{k} < \pa_{2^{n-v+1} - k - 1}$}
		\STATE $ \rhd $ swap($\de_{k} , \de_{2^{n-v+1} - k - 1} $). 
		\STATE $ \rhd $ swap($ \pa_k, \pa_{2^{n-v+1} - k - 1} $) 
		\par
		\hskip \algorithmicindent or $ \de_{k} $ and $ \de_{2^{n-v+1} - k - 1} $.
		\ENDIF 
		\ENDFOR
		\STATE $ \rhd $ Final bias $\epsilon^{increase} \leftarrow \sum\limits_{j=0}^{2^{n-v} - 1} \pa_{j} - $ $ \sum\limits_{j=2^{n-v}}^{2^{n-v+1}} \pa_{j} $
		\STATE Purity increase ratio $\leftarrow$ $ \frac{\epsilon^{increase}}{\epsilon^{in}_v} $
		\ENDWHILE
		\STATE $\epsilon^{f}_{r,v} \leftarrow \epsilon^{increase} $
		\ENDFOR
		\IF{$r == 1$}
		\STATE $\epsilon^{f}_{r,v+1:n} \leftarrow \omega_{v+1:n} $
		\ELSE
		\STATE $\epsilon^{f}_{r,v+1:n} \leftarrow \epsilon^{f}_{r-1,v+1:n} $
		\ENDIF
		\ENDFOR
		\STATE $\mathbf{Print:}$ $\epsilon^{f}$, which displays the limit for each qubit in each round as a $ r \cross n $ matrix.
	\end{algorithmic}
\end{algorithm}	
Figure~\ref{Limits} provides a graphical visualization of the above subroutine.

\subsection{Register Compression Subroutine}  
\label{RCS}
\hspace{0.5cm}
The limits obtained in the previous subroutine serve as targets for \textit{adaptively} initializing the quantum register in each round. This program ensures that all the qubits in the register are exactly initialized till the desired limit in that particular round. This allows us to systematically obtain better cooling in successive rounds. Here we shall account for the reduction in purities of the ancilla qubits within a compression subspace during iteration of the while loop. In effect, we are able to find and count all of the swaps or unitary transformations needed to initialize the quantum register.

\hspace{0.5cm}
Within a given limiting round (line 2), we enter the compression subspace of a particular target qubit ($ x $; line 4) where all the qubits from 1 to $ x-1 $ are disconnected from the subspace because they are purified to the limit corresponding to that round. We then enter the subspace compression subroutine to purify the chosen subspace to its limits corresponding to round $r$. 
\begin{algorithm}[H]
	\caption{\textbf{Register Compression}}
	\label{alg3}
	\begin{algorithmic}[1]
		\REQUIRE Number of qubits, $ n $.
		\REQUIRE Starting biases of the qubits, $ \{ \epsilon^{in}_i \} $.
		\REQUIRE Number of \textit{rounds}; highest possible is $ n - 2 $.
		\REQUIRE Desired \textit{precision}.
		\REQUIRE Round limits (RL) $\leftarrow$ Zero matrix of dimensions $ \textit{rounds} \cross n $
		\REQUIRE \textit{Targets} $\leftarrow$ Limits(n,$ \{ \epsilon^{in}_i \} $,\textit{rounds,precision})
		\ENSURE Sequentially arrive at the cooling limit of all the qubits in the register and count the number of unitary transformation required in the process.
		\STATE $ \rhd $ \textit{complexity} $\leftarrow$ 0
		\FOR{r $\leftarrow$ 1 to \textit{rounds}}
		\STATE $ \rhd $ \textit{Swaps in this round (totswaps)} $\leftarrow$ 0
		\FOR{x $\leftarrow$ 1 to $ n - r - 1 $}
		\STATE $ \rhd $ z $\leftarrow$ 0 
		\STATE $ \rhd $ \textit{Swaps for subspace compression (tttswaps)} $\leftarrow$ 0
		\STATE $ \rhd $ [\textit{tttswaps}, Subspace limits (SL)] $\leftarrow$ SI(n,\textit{rounds},\textit{Targets},x,z,\textit{tttswaps},\textit{precision},RL)
		\STATE $ \rhd $ \textit{totswaps} $\leftarrow$ \textit{totswaps} + \textit{tttswaps}
		\STATE $ \rhd $ RL $\leftarrow$ SL 
		\ENDFOR
		\IF{$r < \textit{rounds}$}
		\STATE $\text{RL}_{r+1} \leftarrow \text{RL}_{r}$
		\ENDIF
		\STATE \textit{complexity} $\leftarrow$ \textit{complexity} + \textit{totswaps}
		\ENDFOR
		\STATE \textbf{Print:} complexity -- Total number of unitaries leading up to the limits.
		\STATE \textbf{Print:} Round limits -- Matrix of dimensions $ \textit{rounds} \cross n $ with the limit of each qubit in every round.
	\end{algorithmic}
\end{algorithm}
Within the subspace compression subroutine, we enter another subspace -- let us call it the subsubspace (line 4 in \ref{alg4}) -- with the $ v^{th} $ qubit as the target. The compressive while loop (line 11) now accounts for the reduction in purities/biases of the ancilla qubits (lines 39-43) due to entropy transfer out of the target. Line 9. ensures that we enter the compression loop only if the bias/purity of the particular target qubit is less than the limit for that particular round, which is loaded from the subroutine Limits\ref{alg2}. Further, we put conditions for termination of the for loop (responsible for the \oss) within the while loop if the bias/purity of the target qubit in any given subsubspace overshoots the database Limits (lines 28, 30, and 32). This ensures that the first qubit within the subsubspace assumes precedence in the order of purification over the rest leading to a systematic descending order of biases in the qubit register after each limiting round. The subsubspace purities are updated (lines 56-58) only when we enter the compression while loop indicated by the value of $w$. This information is then fed into our main database (RL) for processing in the next ((v+1)$^{th}$) iteration of the for loop. 

\hspace{0.5cm}
If the compressive while loop, and therefore the for loop, succeeds (indicated by the difference in  initial and final purities (line 64)), we enter the recursive conditions taking us back to the subspace compression subroutine if we fall short of the target for the corresponding round (line 68 for the first qubit in the target subsubspace and line 73 for the rest). The subspace compression subroutine is repeated till these conditions fail. The output goes to the register compression subroutine as the total number of the unitaries (tttswaps) and the RL database as we move to the next ((x+1)$^{th}$) target subspace in the register compression subroutine.

\begin{algorithm}[H]
	\caption{\textbf{Subspace Compression}}
	\label{alg4}
	\begin{algorithmic}[1]
		\REQUIRE $ \{ \epsilon_i^{in} \} $, RL, $ n $, \textit{rounds} (r), Targets, tttswaps, precision, x, z.
		\ENSURE The subspace spanning 
		the $ x^{th} $ qubit to the $ n^{th} $ qubit is initialized for the round $r$.
		\STATE $ \rhd $ Swaps initializing a single target subspace (ttswaps) $\leftarrow$ 0
		\STATE $ \rhd $ $ \alpha $ $\leftarrow$ RL$_r$
		\STATE $ \rhd $ $a \leftarrow \text{RL}_r $  
		\FOR{v $\leftarrow$ $x$ to $ n - 1 $}
		\STATE $ \rhd $ $s \leftarrow 0$
		\STATE $ \rhd $ Purity increase ratio $\leftarrow$ 0
		\STATE $ \rhd $ $ \beta \leftarrow \alpha_{v:n} $
		\STATE $ \rhd $ Swaps initializing the first qubit within the target subspace (tswaps) $\leftarrow$ 0
		\IF{$\alpha_v <$ \textit{Targets}$_{r,v}$}
		\STATE $w \leftarrow 1$
		\WHILE{Purity increase ratio $\neq 1$ within the desired \textit{precision}}
		\STATE $s \leftarrow s + 1$
		\IF{$s > 1$}
		\STATE $ \beta \leftarrow \gamma $		
		\ENDIF 
		\FOR{$j \leftarrow 0$ to $2^{n-v+1} - 1$}
		\STATE 
		\begin{varwidth}[t]{\linewidth} 
			$\rhd$ $ \de_{j} \leftarrow $ $ (j)_{10} $ converted to its binary equivalent $ (j)_{2} $ composed of $ n $ binary digits by appropriately padding $ 0's $ to the left.
		\end{varwidth}	
		\FOR{$i \leftarrow 1$ to $n$}
		\STATE $\rhd$ $ \epsilon_i^+ \leftarrow (1 + \epsilon_i)/2 $ and $ \epsilon_i^- \leftarrow (1 - \epsilon_i)/2 $.
		\STATE $\rhd$ $\pa_{ij}$ $ \leftarrow $ Assigning $ \epsilon_i^+ $ for $ 0_i $ and $ \epsilon_i^- $
		for $ 1_i $ in $ \de_{j} $.
		\ENDFOR 
		\STATE $\rhd$ $\pa_{j}$ $ \leftarrow $ $\prod_{i} \pas_{ij}$
		\ENDFOR
		\STATE $ \rhd $ $ \gamma \leftarrow \beta $
		\STATE $ \rhd $ $ \gamma' \leftarrow \gamma $
		\STATE $ \rhd $ swap $\leftarrow$ 0
		\FOR{$k \leftarrow 0 : 2^{n-v}$}
		\IF{$z$=0 \& $r>$1 \& $v>$$x$ \& $ \gamma_1 \geq $ \textit{targets}$_{r-1,v}$}
		\STATE break
		\ELSIF{$z$=1 \& $r>$1 \& $ \gamma_1 \geq $ \textit{targets}$_{r-1,v}$}
		\STATE break
		\ELSIF{$z$=0 \& $v$$=$$x$ \& $ \gamma_1 \geq $ \textit{targets}$_{r-1,v}$}
		\STATE break
		\ELSIF{$\pa_{k} < \pa_{2^{n-v+1} - k - 1}$}
		\STATE $ \rhd $ swap($\de_{k} , \de_{2^{n-v+1} - k - 1} $).
		\STATE $ \rhd $ swap $\leftarrow$ swap + 1
		\STATE $ \rhd $ swap($ \pa_{k}, \pa_{2^{n-v+1} - k - 1} $)
		\FOR{$i \leftarrow $ $ 1 $ to $ n - v + 1 $}
		\STATE $ \rhd $ $ c \leftarrow 2^{n-v+1-i} $
		\STATE $ \rhd $ $ \gamma_i $ $\leftarrow$ $ \sum\limits_{j=1}^{c} \pa_{j} - \sum\limits_{j=1 + c}^{2*c} \pa_{j} $
		\FOR{$ m \leftarrow 1 $ to $ 2^{i-1} - 1 $}
		\STATE $ \rhd $ $ \gamma_i \leftarrow \gamma_i + \sum\limits_{j = 1 + (2^m)*c}^{(2^m + 1)*c} \pa_{j} - $ $ \sum\limits_{j = 1 + (2^m + 1)*c}^{(2^m + 2)*c} \pa_{j} $
		\ENDFOR
		\algstore{myalg3}
	\end{algorithmic}
\end{algorithm}

\begin{algorithm}[H]
	\begin{algorithmic}[1]
		\algrestore{myalg3} 
		\IF{$\gamma_i < \epsilon_i^{in}$}
		\STATE $\gamma_i \leftarrow \epsilon_i^{in}$.
		\ENDIF
		\ENDFOR
		\ENDIF
		\ENDFOR
		\STATE $ \rhd $ Purity increase ratio $\leftarrow$ $ \frac{\gamma_1}{\gamma_1'} $
		\STATE $ \rhd $ tswaps $\leftarrow$ tswaps + swaps
		\ENDWHILE
		\ELSE
		\STATE $ w \leftarrow 0 $  
		\ENDIF
		\IF{$ w == 1 $}
		\STATE $ \alpha_{v:n} \leftarrow \gamma_{1:n-v+1} $
		\ENDIF
		\STATE $ \rhd $ RL$_{r,v:n}$ $\leftarrow$ $ \alpha_{v:n} $
		\STATE $ \rhd $ ttswaps $\leftarrow$ ttswaps + tswaps
		\ENDFOR
		\STATE $ \rhd $ $ b \leftarrow \text{RL}_r $
		\STATE $ \rhd $ $ g \leftarrow 0 $
		\IF{$a == b$}
		\STATE $g \leftarrow 1$
		\ENDIF
		\STATE tttswaps $\leftarrow$ tttswaps + ttswaps
		\IF{$ \alpha_x $ $<$ \textit{Targets}$_{r,x}$ \& $ g = 0 $}
		\STATE $ \rhd $ z $\leftarrow$ 0
		\STATE $ \rhd $ [\textit{tttswaps},RL] $\leftarrow$ SI(n,\textit{rounds},\textit{Targets},x,z,\textit{tttswaps},\textit{precision},RL)
		\ENDIF
		\FOR{$ i \leftarrow x+1 : n-1 $}
		\IF{$ r > 1 $ \& $ \alpha_i $ $<$ \textit{Targets}$_{r-1,i}$ \& $ g = 0 $ }
		\STATE $ \rhd $ z $\leftarrow$ 1
		\STATE $ \rhd $ [\textit{tttswaps},RL]$\leftarrow$ SI(n,\textit{rounds},\textit{Targets},i,z,\textit{tttswaps},\textit{precision},RL)
		\ENDIF
		\ENDFOR
		\STATE $ \rhd $ Subroutine outcome 1 $\leftarrow$ tttswaps
		\STATE $ \rhd $ Subroutine outcome 2 $\leftarrow$ RL
	\end{algorithmic}
\end{algorithm}	
Figure~\ref{RIS4q} provides a graphical visualization of the register compression subroutine in case of a 4 qubit register.

\hspace{0.5cm}
We note that just using the limiting swap (implemented with the LIM-Comp~\ref{LIMCOMP}) will converge to the limits in the idealized scenario of relaxation times, but for the case of finite relaxation times for the computation and the reset qubit(s) all the optswaps would be necessary to achieve better cooling in lesser steps as demonstrated before~\cite{PhysRevA.96.012330}. This would be especially pertinent during the first few steps which provide the most significant boost in the biases. Even if only the LIM-Comp is available to be used (say due to limitations of the architecture Hamiltonian), the RCS will continue to determine dynamics of HBAC because it would be necessary to make the appropriate transitions within subspaces for reaching the cooling limit for all qubits in the register. In this case, the checkpoint in line 34 of Subspace Compression~\ref{alg4} should be replaced with just the limiting swap for that subspace and the for loop (line 27) would not be necessary.  

\subsection{Example: 3-qubit register}
We implement the RCS for a 3-qubit register with equal initial biases of $ \epsilon = 0.00001 $. This performs the swap $ \ket{011} \leftrightarrow \ket{100} $ sixteen times to obtain the final initial bias of the first qubit with a precision requirement of $ 0.00001 $. The exact numbers for the final bias is $ 0.199998474100793 \cross 10^{(-4)} $, which is about $ 2 \epsilon $ and matches the theoretical result and is in tune with the experimental results obtained earlier with lesser number of rounds~\cite{ryan2008spin,zaiser2018experimental}. As we decrease the required precision, the complexity (no. of rounds or times the swap is performed) also decreases but only with a small decrease in the final bias that is obtained. For precision requirements of 0.0001, 0.001, 0.01 and 0.1, the final biases obtained are $ 0.199987792948719 \cross 10^{(-4)} $, $ 0.199804687481819 \cross 10^{(-4)} $, $ 0.198437499981940 \cross 10^{(-4)} $ and $ 0.187499999986795 \cross 10^{(-4)} $; and the number of swaps (or rounds/complexity) is $ 13 $, $ 9 $, $ 6 $ and $ 3 $ respectively.

\section{Complexity}
\label{comp}
\hspace{0.5cm}
The classical space and time complexity of implementing the register compression algorithm comprising the subroutines Limits and subspace compression can be analyzed block by block. It may be noted that the algorithm doesn't pose any hindrance if existing computing resources such as workstations with multi-core processors are utilized.

\hspace{0.5cm}
However, the quantum computational complexity comprises the total number of \oss $ $ and the quantum time/depth and space complexity~\cite{watrous2009quantum,bernstein1997quantum,koike2010time} of implementing each \os $ $ using the NB-MaxComp. The gate complexity of the NB-MaxComp can be obtained from the multi-qubit version of the Solovay-Kitaev theorem~\cite{nielsen2002quantum} (the Universality Theorem~\cite{watrous2009quantum}) and the several algorithms using specific generating gate sets for approximating the relevant unitaries to a certain target accuracy~\cite{dawson2005solovay}. A useful result of note in this context is the gate and depth complexity for reversible circuits constructed with NOTs, C--NOTs and 2--CNOTs~\cite{zakablukov2017asymptotic}, which if generalized to k-C--l-NOT (with $k$+$l$$=$$N$) gates would be applicable to the NB-MaxComp. Assuming this quantum time or space complexity to be some $ \nu(N) $, the complexity of implementing register compression on a quantum information processor can be found by counting the number of \oss $ $ required to reach the \tg:
\begin{equation}
	C = O(e^{\lambda N} \nu(N))
\end{equation} 
The exponential scaling of the complexity with the number of qubits can be inferred from the following graphs. In Figure~\ref{COMPLEX1}, we plot the complexity to reach the limit when initial biases are all $0.1$.  
In Figure~\ref{COMPLEX2}, we plot the complexity to reach the limit when initial biases are all $0.00001$.
In Figure~\ref{COMPLEX3}, the complexity is plotted for $n = 5$ against changing initial biases by an order of magnitude.
While we did this to get the big-O complexity, specific cases, for example, where one starts with unequal biases and different targets can be straightforwardly obtained from the classical algorithm.

\hspace{0.5cm}
Additionally, for the LIM-Comp and NB-MaxComp, it would be interesting to minimize the quantum cost~\cite{banerjee2009algorithm,rahman2011two,PhysRevA.52.3457} as done for other reversible logic gates~\cite{saligram2013design,pareek2014new} and to optimize their implementation as done for the Toffoli circuit family~\cite{lanyon2009simplifying,abdessaied2016reversible,miller2011elementary,nam2018automated}.

\section{Discussion}
\label{disc}
\hspace{0.5cm}
In summary, we constructed an explicit heat-bath algorithmic cooling algorithm based on optimal closed-system entropy compression among qubits at each cooling step which asymptotically reaches the HBAC limit of cooling each qubit in a quantum register. In doing so, we were able to account for the reduction in entropy of each qubit while cooling a target qubit within the subspace and thus counted the total number of unitary transformations needed to achieve a desired level of cooling in the register. To reach zero temperature (purity 1), we require infinite precision in the RCS which would take infinite number of steps to reach making this consistent with the third law of thermodynamics~\cite{masanes2017general}. In that context, it would be interesting to find how the temperature scales with the time required to cool as we move away from the idealized scenario to finite relaxation times, since we are able to count the exact number of steps in this operational HBAC protocol (Equation~(7) in Reference~\cite{masanes2017general}). Also, the work cost of implementing the NB-MaxComp specific to particular architectures would help check the relation with final attainable temperature (Equation~(1) and (13) in Reference~\cite{masanes2017general}). In this context, we note that the work cost and efficiency for implementing the two-qubit swap $ \ket{01} \leftrightarrow \ket{10} $ when biases of the two qubits are different (else, it would be moot) has been found~\cite{weimer2008local}. The extension of this result to NB-MaxComp would allow us to compute the engine efficiency per qubit expressions and work cost for the transformations.

\hspace{0.5cm}
The optswaps introduced for the closed-system entropy compression may have implications for quantum data compression~\cite{bennett1995quantum} when the signal states are known~\cite{schumacher1995quantum}. Particularly, lossy data compression with 3 qubit - 2 qubit block coding proposed~\cite{jozsa1994new,vaccaro2003quantum} and implemented~\cite{mitsumori2003experimental} before uses the transformation $ \ket{011} \leftrightarrow \ket{100} $. Optswaps and the NB-MaxComp would allow this to be optimally extended to block sizes of more qubits; particularly we could apply the optswaps on Hilbert spaces of  sequentially decreasing dimension as one decreases the size of the block depending on how much information loss can be tolerated~\cite{datta2013one,langford2002generic}. It would be interesting to compare this method with fixed or variable-length coding schemes which work with unknown but identical signal states using the quantum Schur-Weyl transform~\cite{plesch2010efficient,bacon2006efficient,rozema2014quantum,hayashi2002quantum}. 

\hspace{0.5cm}
The RCS could be modified to include different thermalization methods relying on the architecture used which reportedly exceed the limit allowed by relaxation with a conventional heat bath. Examples are, the $ \Gamma_i $ ($i$ denotes the number of qubits) correlated qubit relaxation operation which is a subroutine of the Nuclear Overhauser Effect (NOE) common to NMR systems in tandem with $ X $-gates~\cite{rodriguez2017heat} (a result which is generalizable to superconducting qubits and ion traps), or use of the $ \beta $-swap operation to cool qubits in a micromaser setting~\cite{alhambra2019heat}. 

\begin{figure}
	\centering
	\includegraphics[width=\linewidth]{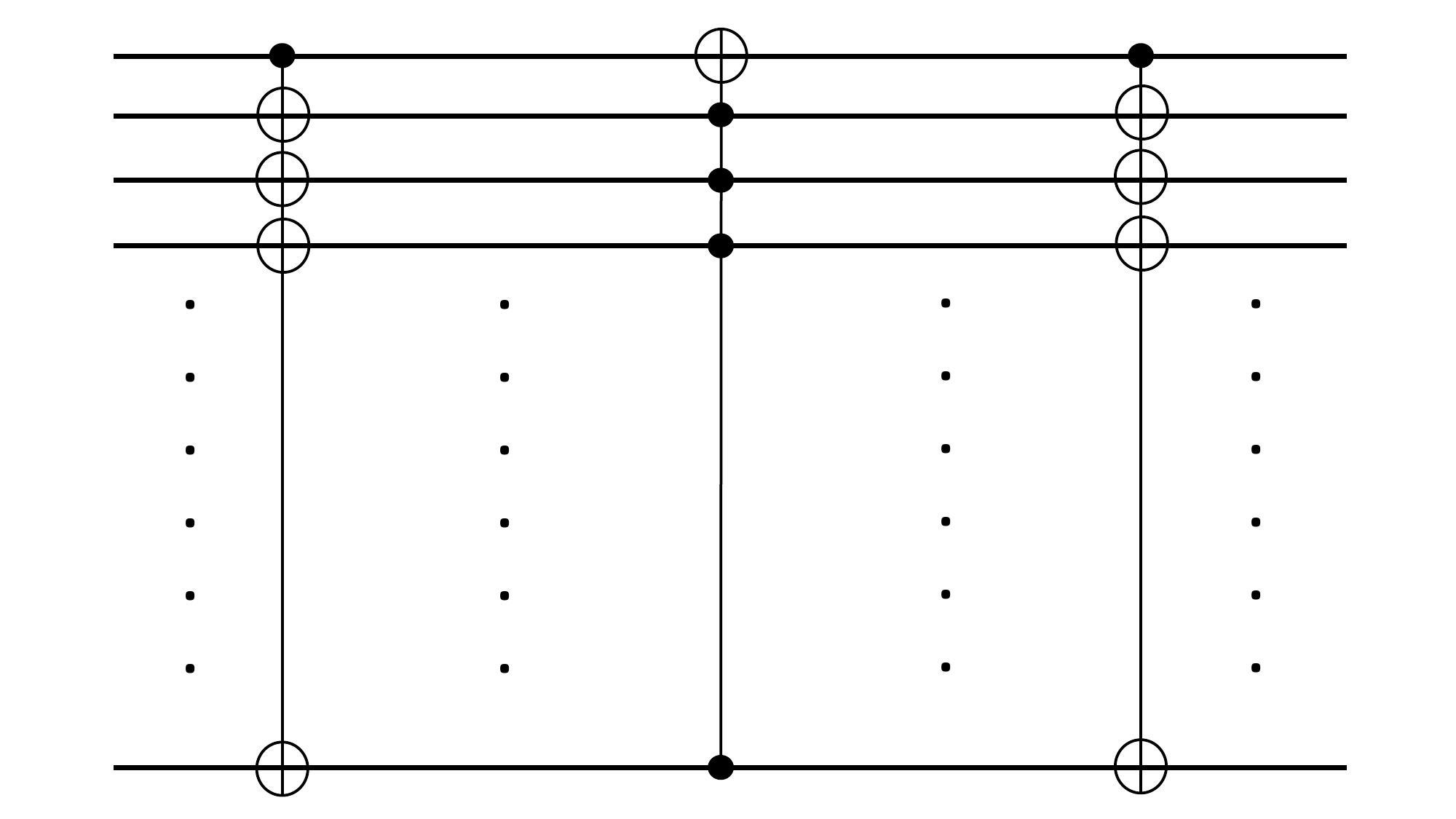}
	\caption{The n-qubit LIM-Comp circuit implements the transformation corresponding to the limiting swap: $ \ket{01111....11} \leftrightarrow \ket{1000....00} $}
	\label{LIMCOMP}
\end{figure}

\begin{figure}
	\centering
	\includegraphics[width=\linewidth]{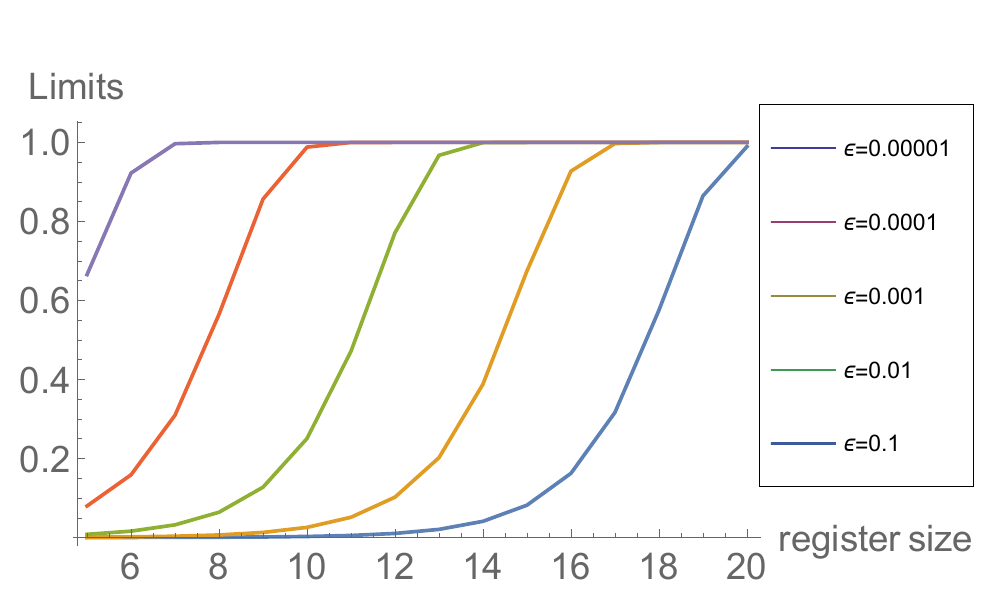}
	\caption{The y-axis represents limits $ \epsilon_{rkn} $ for $ r = n - 2  $ and $ k = 1 $ plotted against the register size $ n $ on the x-axis.}
	\label{Limitsvsn}
\end{figure}

\begin{figure}
	\centering
	\includegraphics[width=\linewidth]{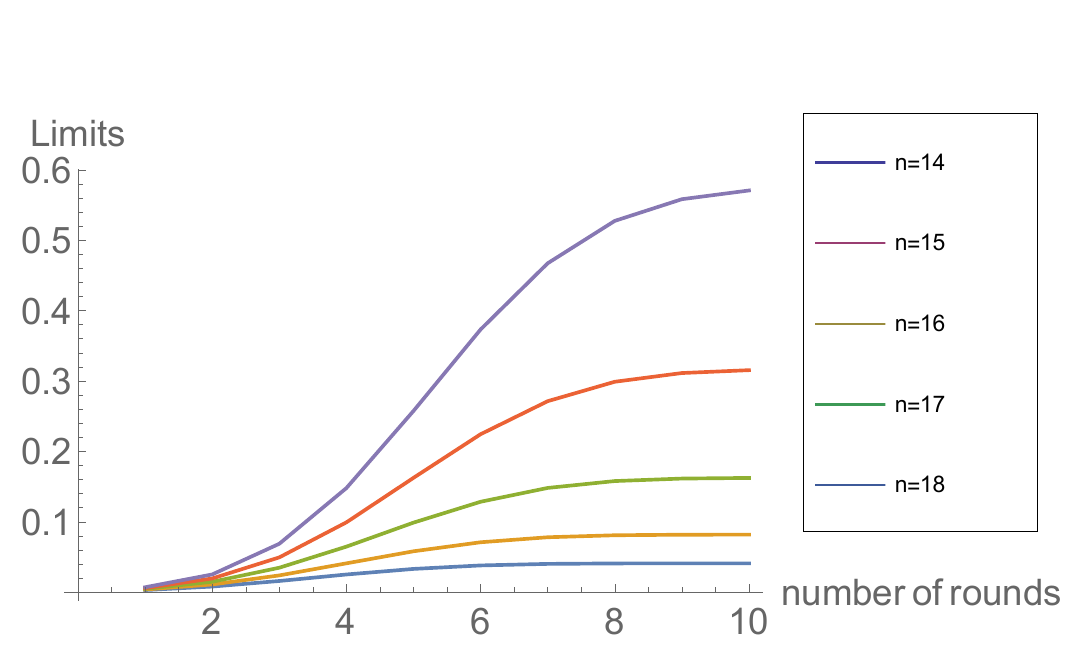}
	\caption{The y-axis represents limits $ \epsilon_{rkn} $ for $ \epsilon = 0.00001  $ and $ k = 1 $ plotted against the number of rounds $ r $ on the x-axis.}
	\label{Limitsvsrlowe}
\end{figure}

\begin{figure}
	\centering
	\includegraphics[width=\linewidth]{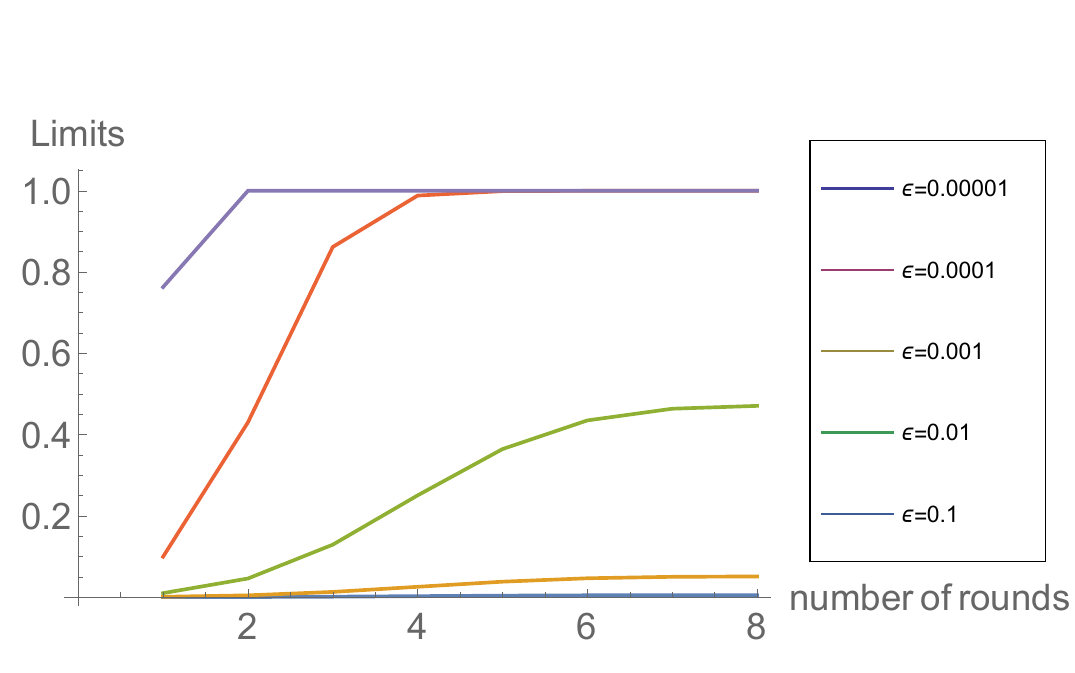}
	\caption{The y-axis represents limits $ \epsilon_{rkn} $ with $ k = 1 $ and $ n = 11 $ for different values of initial biases plotted against the number of rounds $ r $ on the x-axis when register size is $ n = 11 $.}
	\label{fig5}
\end{figure}

\begin{figure}
	\centering
	\includegraphics[width=\linewidth]{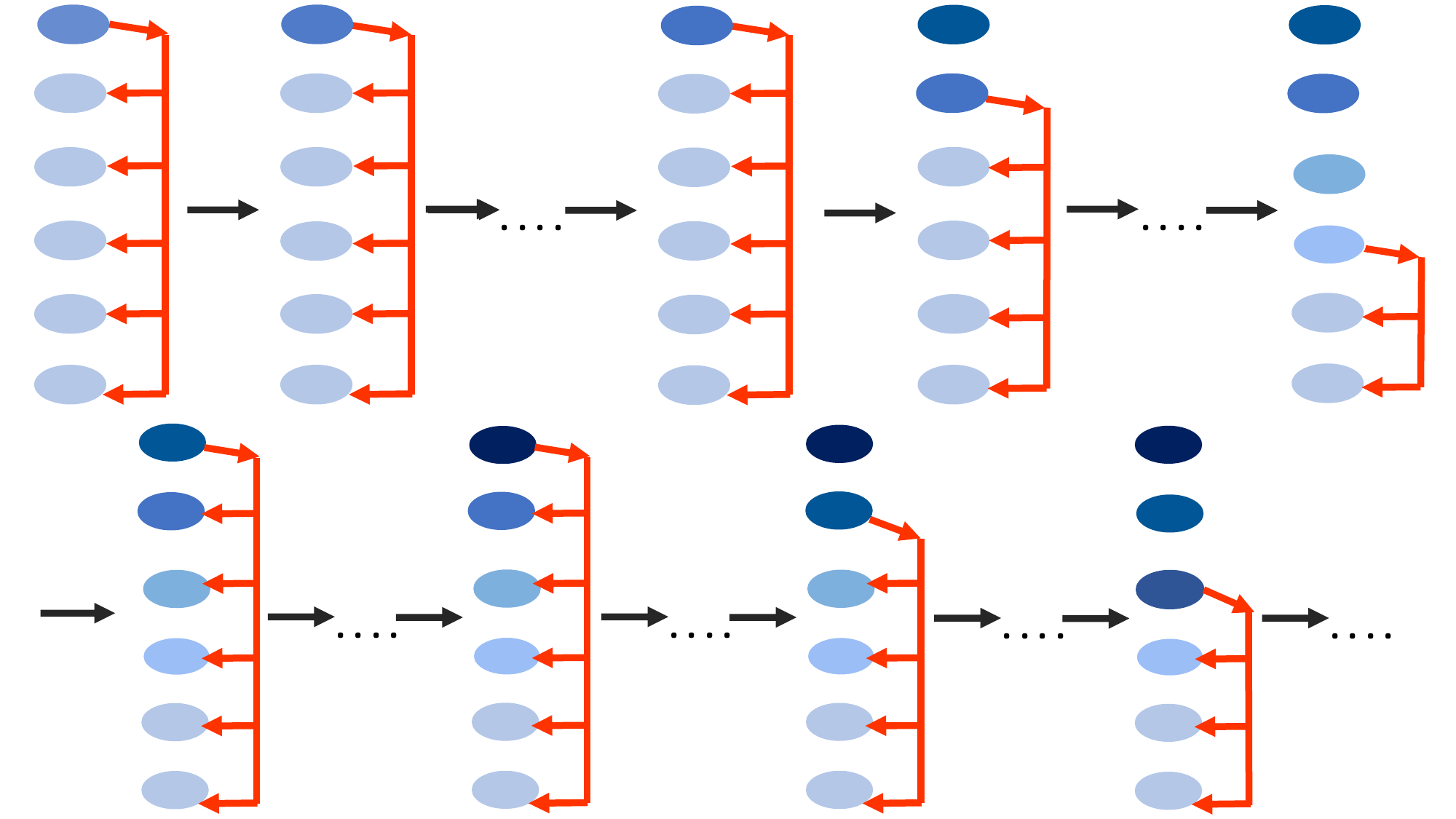}
	\caption{This graphic represents working of the subroutine Limits~\ref{alg2} for a 6-qubit register. Lightest shade of blue represents qubits with default or initial biases. Darker shades of blue represent cooler qubits. 
	Red arrows represent cooling of a particular qubit by transferring its entropy to certain qubits below it. The top half (L-R) of the figure represents purifying the quantum register to round 1 limits of the respective qubits. Bottom half (L-R) represents the use of round 1 limits to purify respective qubits to their round 2 limits.}
	\label{Limits}
\end{figure}

\begin{figure}
	\centering
	\includegraphics[width=\linewidth]{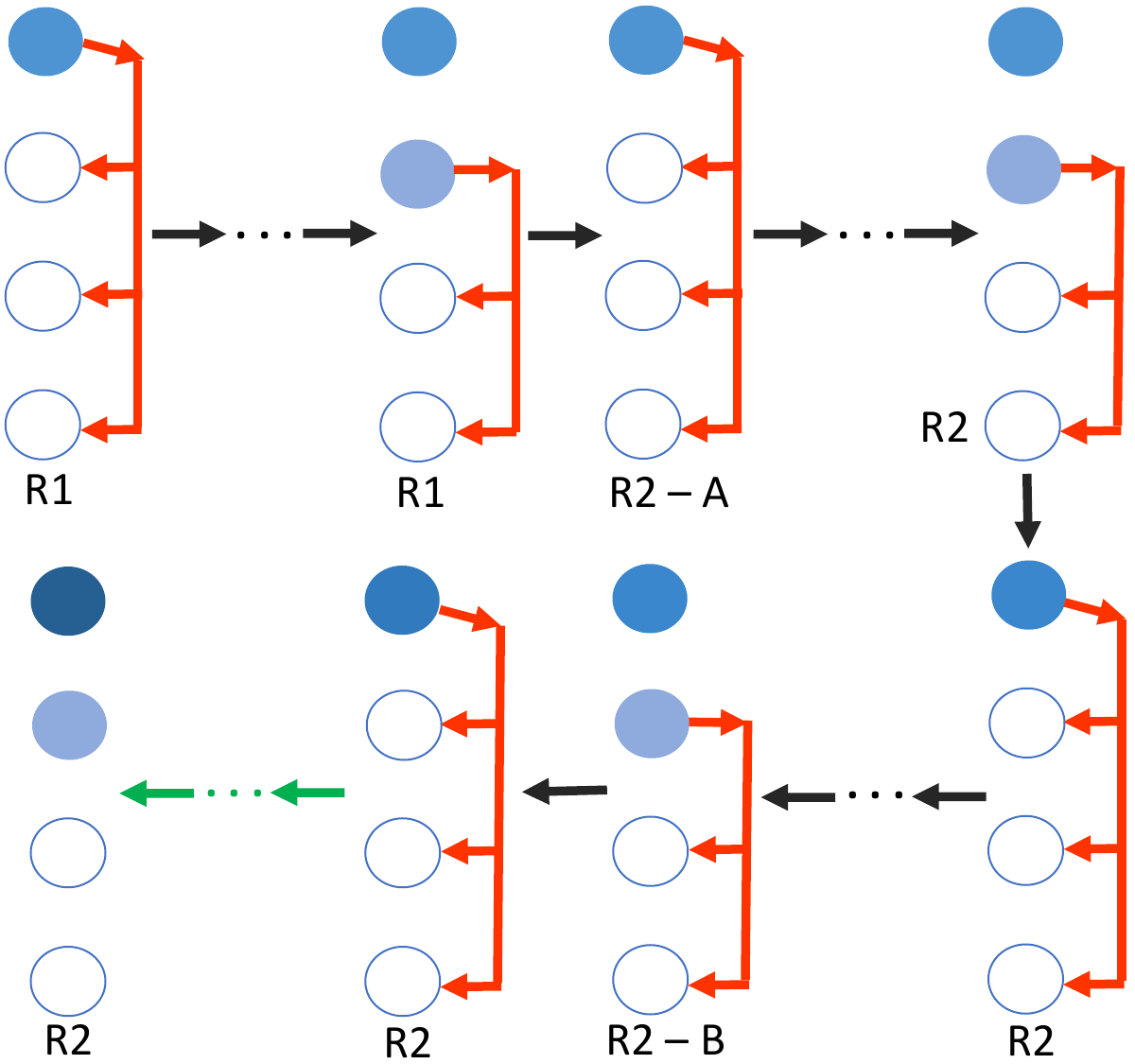}
	\caption{Figure representing working of Register Compression~\ref{alg3} in tandem with Subspace Compression~\ref{alg4} subroutine. The empty circles represent default/initial biases of the qubits.
	Red arrows represent cooling of qubits by transferring its entropy to qubits below it.	R2—A represents the point when subspace initialization subroutine is activated to bring the second qubit back to its round 1 limit. The dashed green arrow represents repeat of steps from R2—A to R2—B till the first qubit is purified to the round 2 limit and the second qubit is purified to the round 1 limit.}
	\label{RIS4q}
\end{figure}

\begin{figure}
	\centering
	\includegraphics[width=\linewidth]{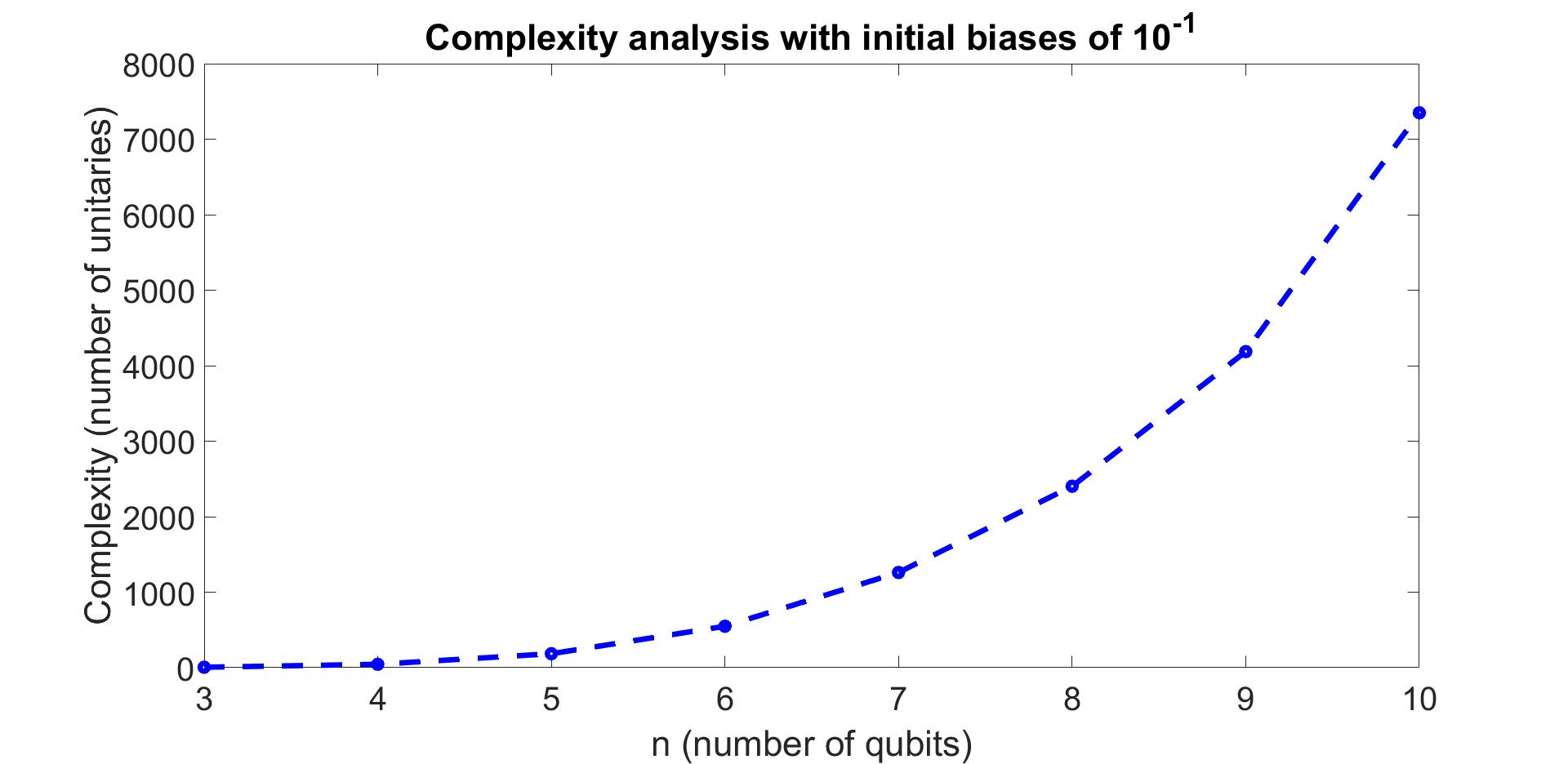}
	\caption{Y-axis represents the number of unitaries it takes to reach the limit of RCS-HBAC for different register sizes (number of qubits) plotted on x-axis for the case where initial biases of all qubits are $ 0.1 $.}
	\label{COMPLEX1}
\end{figure}

\begin{figure}
	\centering
	\includegraphics[width=\linewidth]{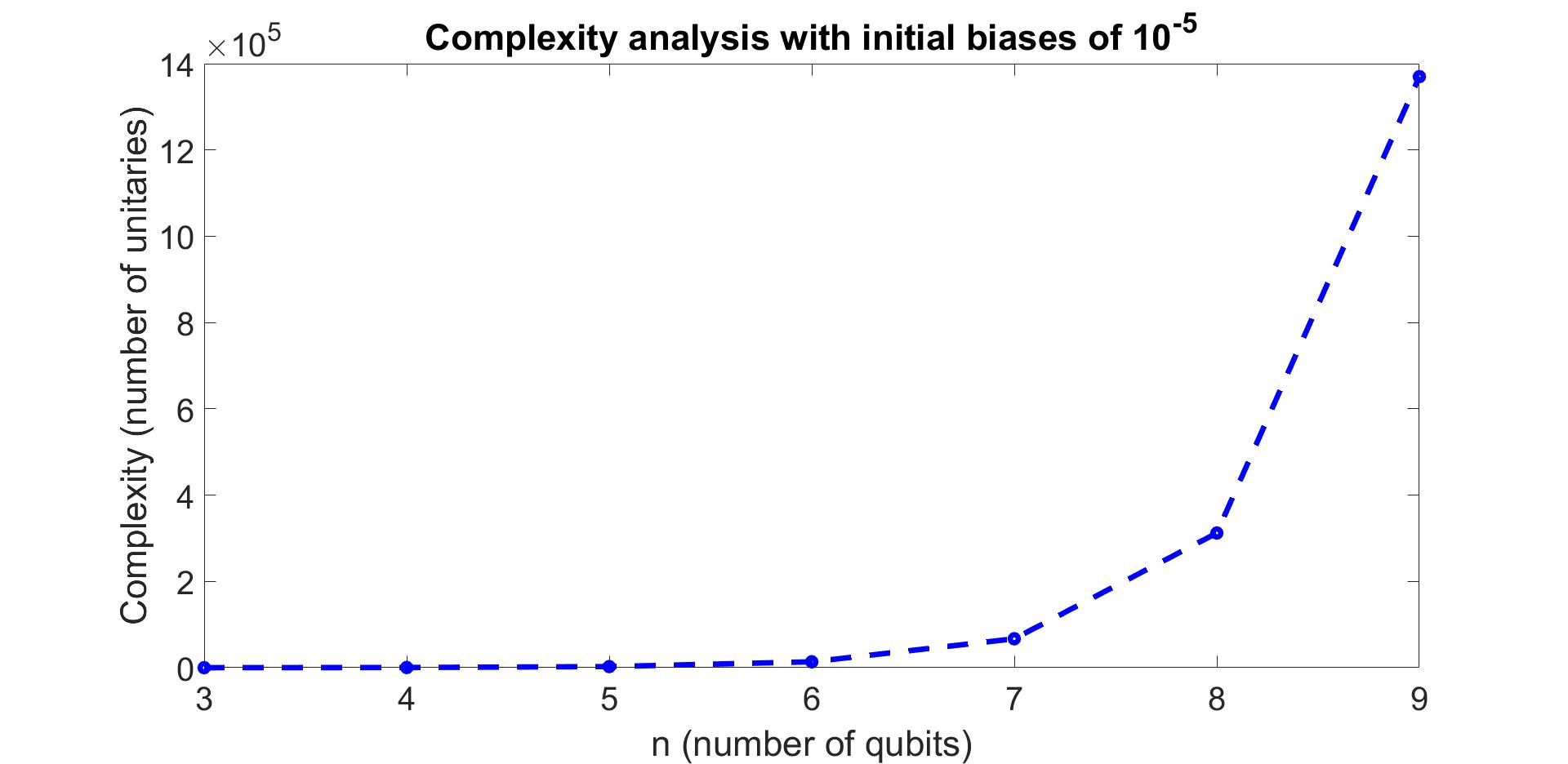}
	\caption{Y-axis represents the number of unitaries it takes to reach the limit of RCS-HBAC for different number of qubits plotted on x-axis for the case where initial biases of all qubits are $ 0.00001 $.}
	\label{COMPLEX2}
\end{figure}

\begin{figure}
	\centering
	\includegraphics[width=\linewidth]{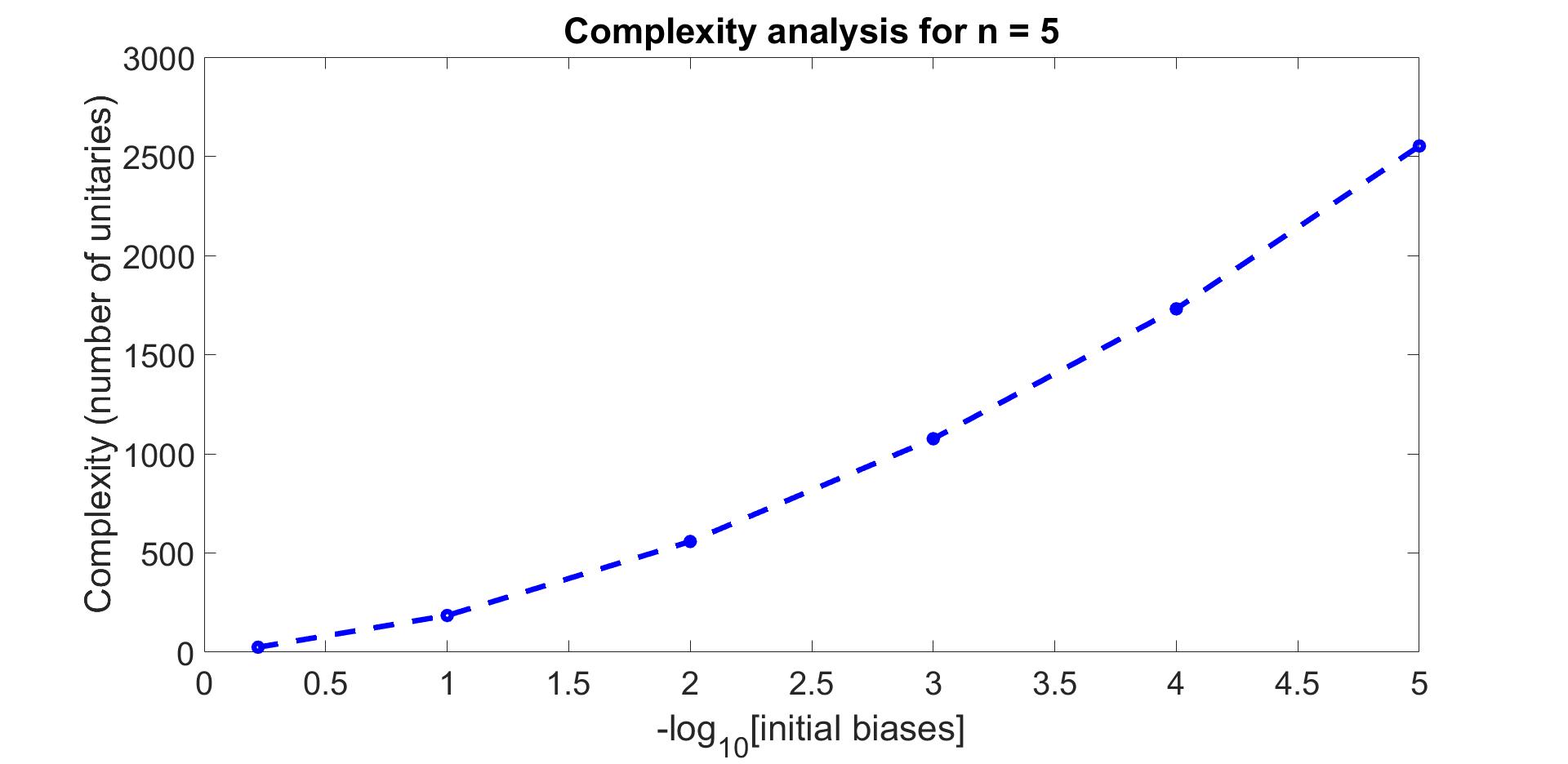}
	\caption{Y-axis represents number of unitaries it takes to reach the limit of RCS-HBAC for cases where initial biases of all qubits are 0.1, 0.01, 0.001, 0.0001, and 0.00001 plotted on a log scale on the x-axis.}
	\label{COMPLEX3}
\end{figure}

\hspace{0.5cm}
Recently~\cite{solfanelli2022quantum}, the Lim-Comp~\ref{LIMCOMP} corresponding to 2 qubits has been used for cooling a superconducting qubit register on IBM's cloud quantum computer (note that the 2-qubit LIM-Comp requires the initial biases to be different), which indicates that HBAC circuits(~\ref{5BMAXCOMP}, \ref{LIMCOMP}) can be used to cool qubits in near-term quantum computers as well.
	
\hspace{0.5cm}
The current HBAC method could be extended to cool a register of qudits (d-level quantum systems), but the definition of optswaps for such systems adds a new layer of complexity. Furthermore, the circuits needed to implement the requisite exchanges in d-level systems would need multi-level NOT and multi-level CONTROL gates, leading to the interesting challenge of defining the ND-MaxComp quantum circuits.

\appendix

\section{Numerical proof examples}
\hspace{0.5cm}
Conjecture~(1) has been numerically verified through Algorithm~\ref{alg1} for the following combinations of initial biases. 
These combinations are chosen to reflect several scenarios which
attempt to find counterexamples that will disprove the conjecture. Specifically, we consider five cases each up to a 19 qubit register and a single case for the 23 qubit register. Verification for larger number of qubits is possible with greater computing power. However, we believe that a counter-example, if it exists, should come through the extreme bias cases we consider rather than higher qubit numbers. These cases are chosen to reflect several scenarios such as randomly selected bias values between (0,1), highly skewed bias values of a few of the qubits, and so forth.
\begin{enumerate}
	\item n = 5: \{0.0147, 0.0893, 0.0881, 0.0448, 0.0737\}, \{0.1405, 0.4176, 0.9066, 0.7843, 0.9499 \}, \{0.0800, 0.00001, 0.00001, 0.1000, 0.1000\}, \{0.8000, 0.0001, 0.0001, 0.0002, 0.0001\}, \{0.00002, 0.9000, 0.1000, 0.2000, 0.0700\}; 
	\item n = 9: \{0.6492, 0.0354, 0.8406, 0.9247, 0.6720, 0.7502, 0.7357, 0.3883, 0.6489\}, \{0.0121, 0.0495, 0.0023, 0.0195, 0.0033, 0.0069, 0.0577, 0.0487, 0.0223\}, \{0.00002, 0.9000, 0.1000, 0.2000, 0.0700, 0.0900, 0.7450, 0.2000,  0.5000\}, \{0.2000, 0.0001, 0.0000, 0.0002, 0.0007, 0.0001, 0.000745, 0.0020, 0.5000\}, \{0.7000, 0.0009, 0.0001, 0.0020, 0.0007, 0.0090, 0.0075, 0.0020, 0.0006\}; 
	\item n = 14: \{0.0665, 0.0025, 0.0308, 0.0268, 0.0536, 0.0557, 0.0132, 0.0343, 0.0313, 0.0453, 0.0497, 0.0529, 0.0194, 0.0476\}, \{0.4586, 0.1139, 0.0833, 0.3489, 0.6718, 0.2383, 0.4097, 0.1567, 0.5259, 0.1786, 0.3542, 0.4894,, 0.6236, 0.6715\}, \{0.7000, 0.0009, 0.0001, 0.0020, 0.000745, 0.0090, 0.0075, 0.0020, 0.0006, 0.0009, 0.00023, 0.0001, 0.0036, 0.0001\}, \{0.0003, 0.0900, 0.0100, 0.0020, 0.0007, 0.0900, 0.0075, 0.0020, 0.0060, 0.0009, 0.0230, 0.7000, 0.0036, 0.0500\}, \{0.0003, 0.0900, 0.1000, 0.2000, 0.0700, 0.9000, 0.7450, 0.2000, 0.6000, 0.0900, 0.2300, 0.7000, 0.3560, 0.5000\};
	\item n = 19: \{0.322, 0.93, 0.212, 0.232, 0.607, 0.569, 0.985, 0.2, 0.1, 0.732, 0.42, 0.35, 0.75, 0.23, 0.369, 0.785, 0.12, 0.21, 0.832\}, \{0.4, 0.000093, 0.0000212, 0.000232, 0.000607, 0.000569, 0.000985, 0.0002, 0.0001, 0.0000732, 0.000042, 0.000035, 0.000075, 0.000023, 0.000369, 0.0000785, 0.00012, 0.00021, 0.0000832\}, \{0.00004, 0.93, 0.212, 0.232, 0.607, 0.569, 0.985, 0.2, 0.1, 0.732, 0.42, 0.35, 0.75, 0.23, 0.369, 0.785, 0.12, 0.21, 0.832\}, \{0.04, 0.00093, 0.00212, 0.00232, 0.0607, 0.569, 0.0985, 0.0002, 0.1, 0.0732, 0.0042, 0.35, 0.075, 0.023, 0.000369, 0.0785, 0.0012, 0.21, 0.00832\}, \{0.000003, 0.000093, 0.0000212, 0.000000232, 0.00000607, 0.0000569, 0.0000985, 0.00002, 0.000001, 0.00000732, 0.000042, 0.000035, 0.000075, 0.0000023, 0.0000369, 0.00000785, 0.0000012, 0.000021, 0.0000832\};
	\item n = 23: \{0.5359, 0.5566, 0.1308, 0.3429, 0.3119, 0.4524, 0.4966, 0.5283, 0.1932, 0.4758, 0.4586, 0.1139, 0.0833, 0.3489, 0.6718, 0.2383, 0.4097, 0.1567, 0.5259, 0.1786, 0.3542, 0.4894, 0.6236\}
\end{enumerate}

\medskip

\medskip
\textbf{Acknowledgements} \par 
\hspace{0.5cm}
Helpful, encouraging remarks from and discussions with Gerardo Adesso, Sougato Bose, Kalyan Chakraborty, Nilanjana Datta, Sebastian Deffner, Azizul Haque, Dintomon Joy, Jason Kestner, Raymond Laflamme, Samuel J.~Lomonaco, T.~S.~Mahesh, Jonathan Oppenheim, Sadegh Raeisi, Aditi Sen De, Ujjwal Sen, Uttam Singh, and Kapil Paranjape and comments on the manuscript from Nayeli A. Rodriguez-Briones are gratefully acknowledged. This work was presented at the ``Internation Symposium on New Frontiers in Quantum Correlations" in SN Bose National Center for Basic Sciences, Kolkata (poster), at ``Quantum Roundabout" in the University of Nottingham (poster), and at the APS Meeting of the Mid-Atlantic section in the University of Maryland, College Park (talk). At different points, the author acknowledges financial support from the DST-INSPIRE fellowship, Govt.~of India, at HRI from the DAE, Govt.~of India, and from the State of Maryland (USA) through UMBC Physics Department.

\begin{figure}
\textbf{Table of Contents}\\
  \centering
  \includegraphics[width=\linewidth, scale=0.1]{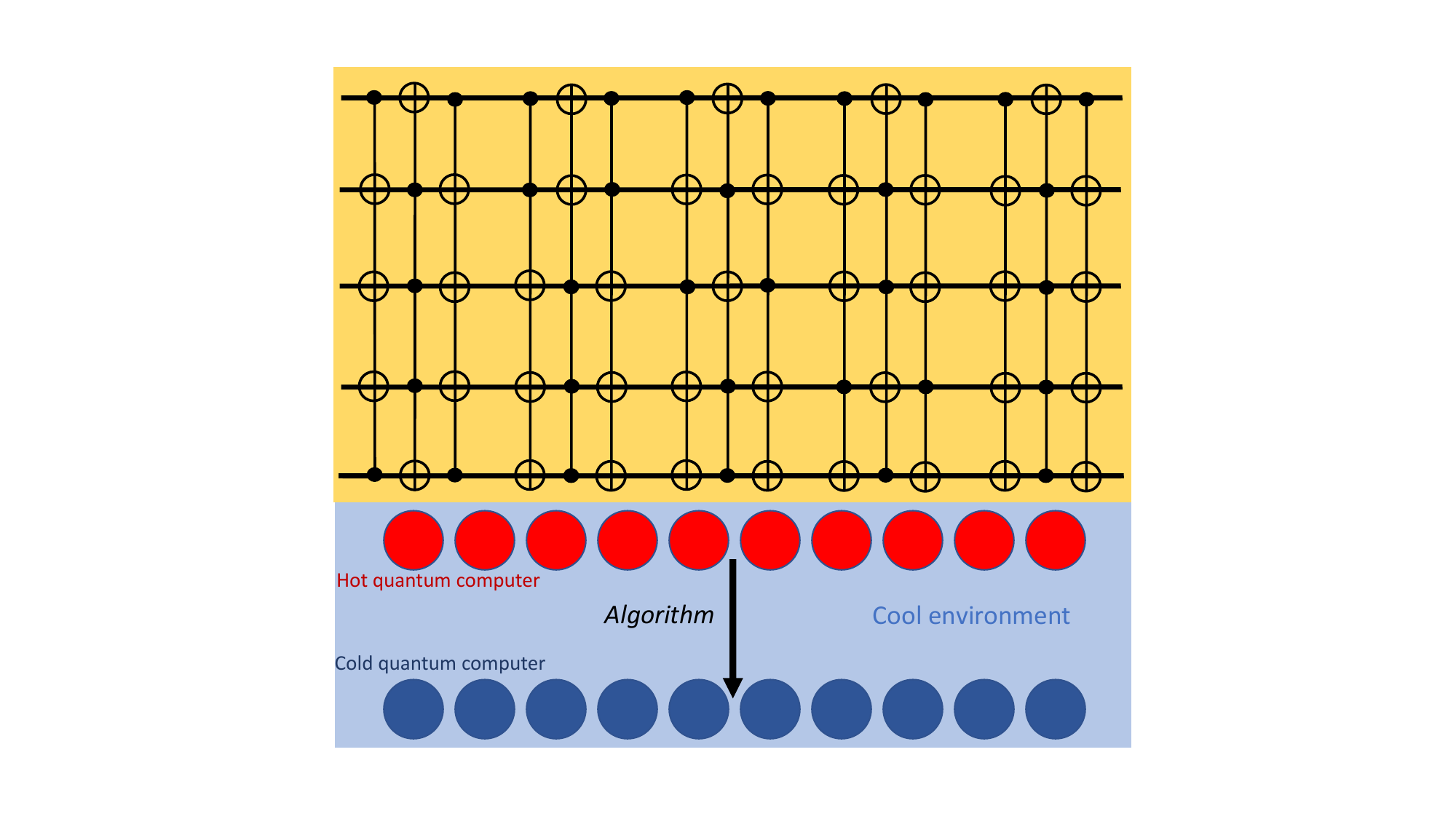}
  \caption*{This work discovers the, as yet unknown, dynamics of heat-bath algorithmic cooling for quantum computers using optimal entropy-compressive unitary transformations that transfer entropy within a multiqubit system (where the qubits could have different initial temperatures) and an operational algorithm for cooling a quantum register. Circuits used for the transformations have implications for quantum data compression and multiqubit quantum thermodynamics.}
\end{figure}


\medskip


\begin{thebibliography}{10}
	\providecommand{\url}[1]{\texttt{#1}}
	\providecommand{\urlprefix}{URL }
	
%
%
%
%
%
%
%
	
	\bibitem{solfanelli2022quantum}
	Andrea Solfanelli, Alessandro Santini, Michele Campisi,
	\newblock \emph{arXiv:2201.13319} \textbf{2022}, \emph{2201.13319}.
	
	\bibitem{divincenzo2000physical}
	D.~P. DiVincenzo,
	\newblock \emph{Fortschritte der Physik} \textbf{2000}, \emph{48} 771.
	
	\bibitem{knill1997theory}
	E.~Knill, R.~Laflamme,
	\newblock \emph{Physical Review A} \textbf{1997}, \emph{55}, 2 900.
	
	\bibitem{PhysRevA.96.012330}
	V.~R. Pande, G.~Bhole, D.~Khurana, T.~S. Mahesh,
	\newblock \emph{Phys. Rev. A} \textbf{2017}, \emph{96} 012330.
	
	\bibitem{boykin2002algorithmic}
	P.~O. Boykin, T.~Mor, V.~Roychowdhury, F.~Vatan, R.~Vrijen,
	\newblock \emph{Proceedings of the National Academy of Sciences} \textbf{2002},
	\emph{99}, 6 3388.
	
	\bibitem{atia2016algorithmic}
	Y.~Atia, Y.~Elias, T.~Mor, Y.~Weinstein,
	\newblock \emph{Physical Review A} \textbf{2016}, \emph{93}, 1 012325.
	
	\bibitem{sorensen1989polarization}
	O.~W. S{\o}rensen,
	\newblock \emph{Progress in Nuclear Magnetic Resonance Spectroscopy}
	\textbf{1989}, \emph{21}, 6 503.
	
	\bibitem{sorensen1990universal}
	O.~W. S{\o}rensen,
	\newblock \emph{Journal of Magnetic Resonance (1969)} \textbf{1990}, \emph{86},
	2 435.
	
	\bibitem{sorensen1991entropy}
	O.~W. S{\o}rensen,
	\newblock \emph{Journal of Magnetic Resonance (1969)} \textbf{1991}, \emph{93},
	3 648.
	
	\bibitem{schulman1999molecular}
	L.~J. Schulman, U.~V. Vazirani,
	\newblock In \emph{Proceedings of the thirty-first annual ACM symposium on
		Theory of computing}. ACM, \textbf{1999} 322--329.
	
	\bibitem{von195113}
	J.~Von~Neumann,
	\newblock \emph{Appl. Math Ser} \textbf{1951}, \emph{12}, 36-38 5.
	
	\bibitem{baugh2005experimental}
	J.~Baugh, O.~Moussa, C.~Ryan, A.~Nayak, R.~Laflamme,
	\newblock \emph{Nature} \textbf{2005}, \emph{438}, 7067 470.
	
	\bibitem{brassard2005experimental}
	G.~Brassard, Y.~Elias, J.~M. Fernandez, H.~Gilboa, J.~A. Jones, T.~Mor,
	Y.~Weinstein, L.~Xiao,
	\newblock \emph{arXiv preprint quant-ph/0511156} \textbf{2005}.
	
	\bibitem{elias2006optimal}
	Y.~Elias, J.~M. Fernandez, T.~Mor, Y.~Weinstein,
	\newblock \emph{Israel Journal of Chemistry} \textbf{2006}, \emph{46}, 4 371.
	
	\bibitem{mor2005algorithmic}
	T.~Mor, V.~Roychowdhury, S.~Lloyd, J.~M. Fernandez, Y.~Weinstein,
	\newblock Algorithmic cooling, \textbf{2005},
	\newblock US Patent 6,873,154.
	
	\bibitem{park2015hyperfine}
	D.~K. Park, G.~Feng, R.~Rahimi, S.~Labruy{\`e}re, T.~Shibata, S.~Nakazawa,
	K.~Sato, T.~Takui, R.~Laflamme, J.~Baugh,
	\newblock \emph{Quantum Information Processing} \textbf{2015}, \emph{14}, 7
	2435.
	
	\bibitem{park2016heat}
	D.~K. Park, N.~A. Rodriguez-Briones, G.~Feng, R.~Rahimi, J.~Baugh, R.~Laflamme,
	\newblock In \emph{Electron Spin Resonance (ESR) Based Quantum Computing},
	227--255. Springer, \textbf{2016}.
	
	\bibitem{kaye2007cooling}
	P.~Kaye,
	\newblock \emph{Quantum Information Processing} \textbf{2007}, \emph{6}, 4 295.
	
	\bibitem{ryan2008spin}
	C.~Ryan, O.~Moussa, J.~Baugh, R.~Laflamme,
	\newblock \emph{Physical review letters} \textbf{2008}, \emph{100}, 14 140501.
	
	\bibitem{elias2011heat}
	Y.~Elias, H.~Gilboa, T.~Mor, Y.~Weinstein,
	\newblock \emph{Chemical Physics Letters} \textbf{2011}, \emph{517}, 4-6 126.
	
	\bibitem{zaiser2018experimental}
	S.~Zaiser, B.~Masatth, D.~Rao, S.~Raeisi, J.~Wrachtrup,
	\newblock \emph{arXiv preprint arXiv:1812.06252} \textbf{2018}.
	
	\bibitem{atia2014quantum}
	Y.~Atia, Y.~Elias, T.~Mor, Y.~Weinstein,
	\newblock \emph{International Journal of Quantum Information} \textbf{2014},
	\emph{12}, 05 1450031.
	
	\bibitem{mor:2008}
	T.~Mor,
	\newblock In M.-Y. Kao, editor, \emph{Encyclopedia of Algorithms}, 30--37.
	Springer US, New York, \textbf{2008}.
	
	\bibitem{chang2001nmr}
	D.~E. Chang, L.~M. Vandersypen, M.~Steffen,
	\newblock \emph{Chemical physics letters} \textbf{2001}, \emph{338}, 4 337.
	
	\bibitem{fernandez2004algorithmic}
	J.~M. Fernandez, S.~Lloyd, T.~Mor, V.~Roychowdhury,
	\newblock \emph{International Journal of Quantum Information} \textbf{2004},
	\emph{2}, 04 461.
	
	\bibitem{brassard2014prospects}
	G.~Brassard, Y.~Elias, T.~Mor, Y.~Weinstein,
	\newblock \emph{The European Physical Journal Plus} \textbf{2014}, \emph{129},
	11 258.
	
	\bibitem{elias2007optimal}
	Y.~Elias, J.~M. Fernandez, T.~Mor, Y.~Weinstein,
	\newblock In \emph{International Conference on Unconventional Computation}.
	Springer, \textbf{2007} 2--26.
	
	\bibitem{elias2011semioptimal}
	Y.~Elias, T.~Mor, Y.~Weinstein,
	\newblock \emph{Physical Review A} \textbf{2011}, \emph{83}, 4 042340.
	
	\bibitem{schulman2005physical}
	L.~J. Schulman, T.~Mor, Y.~Weinstein,
	\newblock \emph{Physical review letters} \textbf{2005}, \emph{94}, 12 120501.
	
	\bibitem{schulman2007physical}
	L.~J. Schulman, T.~Mor, Y.~Weinstein,
	\newblock \emph{SIAM Journal on Computing} \textbf{2007}, \emph{36}, 6 1729.
	
	\bibitem{raeisi2015asymptotic}
	S.~Raeisi, M.~Mosca,
	\newblock \emph{Physical review letters} \textbf{2015}, \emph{114}, 10 100404.
	
	\bibitem{rodriguez2016achievable}
	N.~A. Rodr{\'\i}guez-Briones, R.~Laflamme,
	\newblock \emph{Physical review letters} \textbf{2016}, \emph{116}, 17 170501.
	
	\bibitem{raeisi2019novel}
	S.~Raeisi, M.~Kieferov{\'a}, M.~Mosca,
	\newblock \emph{Physical Review Letters} \textbf{2019}, \emph{122}, 22 220501.
	
	\bibitem{nielsen2002quantum}
	M.~A. Nielsen, I.~Chuang,
	\newblock Quantum computation and quantum information, \textbf{2002}.
	
	\bibitem{watrous2009quantum}
	J.~Watrous,
	\newblock \emph{Encyclopedia of complexity and systems science} \textbf{2009},
	7174--7201.
	
	\bibitem{bernstein1997quantum}
	E.~Bernstein, U.~Vazirani,
	\newblock \emph{SIAM Journal on computing} \textbf{1997}, \emph{26}, 5 1411.
	
	\bibitem{koike2010time}
	T.~Koike, Y.~Okudaira,
	\newblock \emph{Physical Review A} \textbf{2010}, \emph{82}, 4 042305.
	
	\bibitem{dawson2005solovay}
	C.~M. Dawson, M.~A. Nielsen,
	\newblock \emph{arXiv preprint quant-ph/0505030} \textbf{2005}.
	
	\bibitem{zakablukov2017asymptotic}
	D.~V. Zakablukov,
	\newblock \emph{Journal of Computer and System Sciences} \textbf{2017},
	\emph{84} 132.
	
	\bibitem{banerjee2009algorithm}
	A.~Banerjee, A.~Pathak,
	\newblock \emph{arXiv preprint arXiv:0910.2129} \textbf{2009}.
	
	\bibitem{rahman2011two}
	M.~M. Rahman, A.~Banerjee, G.~W. Dueck, A.~Pathak,
	\newblock In \emph{2011 41st IEEE International Symposium on Multiple-Valued
		Logic}. IEEE, \textbf{2011} 86--92.
	
	\bibitem{PhysRevA.52.3457}
	A.~Barenco, C.~H. Bennett, R.~Cleve, D.~P. DiVincenzo, N.~Margolus, P.~Shor,
	T.~Sleator, J.~A. Smolin, H.~Weinfurter,
	\newblock \emph{Phys. Rev. A} \textbf{1995}, \emph{52} 3457.
	
	\bibitem{saligram2013design}
	R.~Saligram, S.~S. Hegde, S.~A. Kulkarni, H.~Bhagyalakshmi, M.~Venkatesha,
	\newblock \emph{International Journal of VLSI Design \& Communication Systems}
	\textbf{2013}, \emph{4}, 3 53.
	
	\bibitem{pareek2014new}
	V.~Pareek,
	\newblock \emph{arXiv preprint arXiv:1410.2373} \textbf{2014}.
	
	\bibitem{lanyon2009simplifying}
	B.~P. Lanyon, M.~Barbieri, M.~P. Almeida, T.~Jennewein, T.~C. Ralph, K.~J.
	Resch, G.~J. Pryde, J.~L. O’brien, A.~Gilchrist, A.~G. White,
	\newblock \emph{Nature Physics} \textbf{2009}, \emph{5}, 2 134.
	
	\bibitem{abdessaied2016reversible}
	R.~D.~a. Nabila~Abdessaied,
	\newblock \emph{Reversible and Quantum Circuits: Optimization and Complexity
		Analysis},
	\newblock Springer International Publishing, 1 edition, \textbf{2016}.
	
	\bibitem{miller2011elementary}
	D.~M. Miller, R.~Wille, Z.~Sasanian,
	\newblock In \emph{2011 41st IEEE International Symposium on Multiple-Valued
		Logic}. IEEE, \textbf{2011} 288--293.
	
	\bibitem{nam2018automated}
	Y.~Nam, N.~J. Ross, Y.~Su, A.~M. Childs, D.~Maslov,
	\newblock \emph{npj Quantum Information} \textbf{2018}, \emph{4}, 1 23.
	
	\bibitem{masanes2017general}
	L.~Masanes, J.~Oppenheim,
	\newblock \emph{Nature communications} \textbf{2017}, \emph{8} 14538.
	
	\bibitem{weimer2008local}
	H.~Weimer, M.~J. Henrich, F.~Rempp, H.~Schr{\"o}der, G.~Mahler,
	\newblock \emph{EPL (Europhysics Letters)} \textbf{2008}, \emph{83}, 3 30008.
	
	\bibitem{bennett1995quantum}
	C.~H. Bennett,
	\newblock \emph{Physics Today} \textbf{1995}, \emph{48}, 10 24.
	
	\bibitem{schumacher1995quantum}
	B.~Schumacher,
	\newblock \emph{Physical Review A} \textbf{1995}, \emph{51}, 4 2738.
	
	\bibitem{jozsa1994new}
	R.~Jozsa, B.~Schumacher,
	\newblock \emph{Journal of Modern Optics} \textbf{1994}, \emph{41}, 12 2343.
	
	\bibitem{vaccaro2003quantum}
	J.~A. Vaccaro, Y.~Mitsumori, S.~M. Barnett, E.~Andersson, A.~Hasegawa,
	M.~Takeoka, M.~Sasaki,
	\newblock In \emph{International Symposium on Stochastic Algorithms}. Springer,
	\textbf{2003} 98--107.
	
	\bibitem{mitsumori2003experimental}
	Y.~Mitsumori, J.~A. Vaccaro, S.~M. Barnett, E.~Andersson, A.~Hasegawa,
	M.~Takeoka, M.~Sasaki,
	\newblock \emph{Physical review letters} \textbf{2003}, \emph{91}, 21 217902.
	
	\bibitem{datta2013one}
	N.~Datta, J.~M. Renes, R.~Renner, M.~M. Wilde,
	\newblock \emph{IEEE Transactions on Information Theory} \textbf{2013},
	\emph{59}, 12 8057.
	
	\bibitem{langford2002generic}
	J.~Langford,
	\newblock \emph{Physical Review A} \textbf{2002}, \emph{65}, 5 052312.
	
	\bibitem{plesch2010efficient}
	M.~Plesch, V.~Bu{\v{z}}ek,
	\newblock \emph{Physical Review A} \textbf{2010}, \emph{81}, 3 032317.
	
	\bibitem{bacon2006efficient}
	D.~Bacon, I.~L. Chuang, A.~W. Harrow,
	\newblock \emph{Physical review letters} \textbf{2006}, \emph{97}, 17 170502.
	
	\bibitem{rozema2014quantum}
	L.~A. Rozema, D.~H. Mahler, A.~Hayat, P.~S. Turner, A.~M. Steinberg,
	\newblock \emph{Physical review letters} \textbf{2014}, \emph{113}, 16 160504.
	
	\bibitem{hayashi2002quantum}
	M.~Hayashi, K.~Matsumoto,
	\newblock \emph{Physical Review A} \textbf{2002}, \emph{66}, 2 022311.
	
	\bibitem{rodriguez2017heat}
	N.~A. Rodriguez-Briones, J.~Li, X.~Peng, T.~Mor, Y.~Weinstein, R.~Laflamme,
	\newblock \emph{New Journal of Physics} \textbf{2017}, \emph{19}, 11 113047.
	
	\bibitem{alhambra2019heat}
	{\'A}.~M. Alhambra, M.~Lostaglio, C.~Perry,
	\newblock \emph{Quantum} \textbf{2019}, \emph{3} 188.
	
\end{thebibliography}
\end{document}